\documentclass[prd,twocolumn,superscriptaddress,floatfix,nofootinbib]{revtex4-2}

\pdfoutput=1 
\usepackage[T1]{fontenc} 
\usepackage{amsmath,amssymb,amsfonts,amsthm}
\usepackage{graphicx}
\usepackage[usenames,dvipsnames]{color}
\usepackage{enumitem}
\usepackage{verbatim}
\usepackage[percent]{overpic}
\usepackage{rotating}
\usepackage[colorlinks=true]{hyperref}
\usepackage{array}
\usepackage{mathtools}
\usepackage{lmodern}
\usepackage[normalem]{ulem}
\usepackage{braket}
\usepackage{tensor}
\usepackage{dsfont}
\usepackage{bm}
\usepackage{tikz}

\usepackage{mathrsfs}

\usetikzlibrary{decorations.pathmorphing}
\usepackage{CJKutf8}

\definecolor{patriarch}{rgb}{0.5, 0.0, 0.5}
\definecolor{pigmentgreen}{rgb}{0.0, 0.65, 0.31}
\definecolor{darkpastelgreen}{rgb}{0.01, 0.75, 0.24}
\definecolor{darkraspberry}{rgb}{0.53, 0.15, 0.34}

\usetikzlibrary{positioning}
\usetikzlibrary{calc,through,backgrounds}

\theoremstyle{definition}
\newtheorem{definition}{Definition}[section]
\newtheorem{theorem}{Theorem}[section]

\allowdisplaybreaks

\DeclareMathOperator{\sgn}{sgn}
\DeclareMathOperator{\arccosh}{arccosh}

\newcommand{\sx}{\mathsf{x}}

\newcommand{\ii}{\mathsf{i}}

\definecolor{patriarch}{rgb}{0.5, 0.0, 0.5}

\begin{document}

\title{Deep in the knotted black hole}

\author{Massimiliano Spadafora}
\email{mspadafo@sissa.it}

\affiliation{SISSA, Via Bonomea 265, 34136 Trieste, Italy}

\author{Manar Naeem}
\email{manar.naeem@uwaterloo.ca}
\affiliation{Department of Physics and Astronomy, University of Waterloo, Waterloo, Ontario, N2L 3G1, Canada}
\affiliation{Institute for Quantum Computing, University of Waterloo, Waterloo, Ontario, N2L 3G1, Canada}

\author{María R. Preciado-Rivas}
\email{mrpreciadorivas@uwaterloo.ca}
\affiliation{Institute for Quantum Computing, University of Waterloo, Waterloo, Ontario, N2L 3G1, Canada}
\affiliation{Department of Applied Mathematics, University of Waterloo, Waterloo, Ontario, N2L 3G1, Canada}
\affiliation{Waterloo Centre for Astrophysics, University of Waterloo, Waterloo, ON, N2L 3G1, Canada}

\author{Robert B. Mann}
\email{rbmann@uwaterloo.ca}
\affiliation{Department of Physics and Astronomy, University of Waterloo, Waterloo, Ontario, N2L 3G1, Canada}
\affiliation{Institute for Quantum Computing, University of Waterloo, Waterloo, Ontario, N2L 3G1, Canada}
\affiliation{Perimeter Institute for Theoretical Physics,  Waterloo, Ontario, N2L 2Y5, Canada}

\author{Jorma Louko}
\email{jorma.louko@nottingham.ac.uk}
\affiliation{School of Mathematical Sciences, University of Nottingham, Nottingham NG7 2RD, UK}
\date{December 2024; revised March 2025.\\ aaPublished in Phys.\ Rev.\ D \textbf{111}, 065013 (2025), doi.org/10.1103/PhysRevD.111.065013.\\ aaFor Open Access purposes, this Author Accepted Manuscript is made available under CC BY public copyright.}
\begin{abstract}
We consider the transition rate of a freely falling Unruh-DeWitt detector, coupled linearly to  a massless scalar quantum field prepared in the Hartle-Hawking-Israel state, as a probe of the interior of a black hole.  
Specifically, we consider the transition rate of a detector in the spinless 
Ba\~nados-Teitelboim-Zanelli (BTZ) black hole as it freely falls toward and across the horizon and compare it to the corresponding situation for an $\mathbb{R}\text{P}^{2}$ geon. 
Both the BTZ black hole and its geon counterpart are quotients of $\text{AdS}_3$ spacetime that are identical exterior to the horizon but have different interior topologies.  We find outside the horizon that the rates are qualitatively similar, but with the amplitude in the geon spacetime larger than in the BTZ case. Once the detector crosses the horizon, there are notable distinctions characterized by different discontinuities in the temporal derivative of the response rate. These discontinuities can appear outside the horizon if the detector is switched on at a sufficiently early time,  within the past white hole horizon.  In general, the detector can act as an `early warning system'  that both spots the black hole horizon and discerns its interior topology.
\end{abstract}

\maketitle

\section{Introduction}

The cosmic censorship conjecture \cite{Penrose:1999vj} states that curvature singularities that form in the future of an initial spacelike surface having regular initial conditions must be ``hidden'' inside event horizons, making them effectively inaccessible to any classical observer.  An extension of this is the topological censorship theorem,
which asserts that any topological structure collapses too quickly to allow light to traverse it, thus forbidding
distant observers to probe the topology of spacetime~\cite{Friedman_1995}. Consequently, all nontrivial topological structures (such as wormholes, handles, crosscaps, etc.) must be concealed behind the horizon of a black hole \cite{Galloway:1996hx,Galloway:1999bp}. This conjecture explains an apparent paradox where the actual topology of spacetime seems trivial, whereas general relativity allows for any type of topology.

When quantum fields on the spacetime are considered,  
the situation is notably different.  
It has been shown that local observers equipped with 
quantum detectors idealized as two-level  systems, or qubits, also known as Unruh-DeWitt (UDW) detectors \cite{Unruh_effect,DeWitt1979}, can
probe the global topology of a spacetime 
when the field has been prepared in a state that knows about the global topology~\cite{Louko:1998dj,Smith_2014}. Specifically, the transition rate of a static UDW detector outside a black hole is sensitive to the interior topology of the black hole, even if the metric exterior to its horizon is identical to that of a topologically trivial black hole
(for which the transition rate of
a static detector thermalizes at the Hawking temperature~\cite{Hodgkinson:2012mr,Hodgkinson:2014iua}).
Black hole spacetimes that have only one exterior region, known as geon black holes~\cite{Friedman_1995}, have played a useful role in understanding Hawking radiation~\cite{Louko:1998dj}, the AdS-CFT correspondence conjecture \cite{Louko:1998hc,Louko:2000tp,Galloway:1999br}, and vacuum entanglement \cite{Martin-Martinez:2015qwa,Henderson:2022oyd}.

There has been growing interest in the response of detectors that freely fall into a black hole and cross its horizon. There appears to be a general expectation 
(perhaps rooted in the equivalence principle)
that such a detector will not thermalize, but characterising the detector's response more precisely is a topic of continuing study. 
An adiabatic formalism for estimating the detector's effective temperature has been developed in \cite{Barbado:2011dx,Barcelo:2010pj,Barbado:2012pt,Smerlak:2013sga,Barbado:2016nfy}.
A~detector freely falling through a stationary cavity in a ($3+1$)-dimensional Schwarzschild background 
was found to have a different response from that of an equivalently accelerated detector travelling through an inertial cavity in flat spacetime~\cite{Ahmadzadegan:2013iua}. 
A~detector falling freely toward and across the horizon of a Schwarzschild black hole has been analysed numerically in \cite{Ng2022,Shallue:2025zto}. 
An operational definition for an effective temperature that accommodates the brief time that the detector spends near the horizon was proposed in~\cite{Shallue:2025zto}. 

Investigation of the response of an infalling detector is technically quite formidable, and so other studies have
resorted to simpler lower-dimensional settings, 
such as the $(1+1)$-dimensional Schwarzschild black hole, obtained by neglecting the angular part of the standard $(3+1)$-dimensional Schwarzschild metric. Taking the detector to be coupled to the momentum of a scalar field, it has been shown in this setting that the thermal character of the  transition rate of a  freely falling UDW detector is gradually lost \cite{Juarez-Aubry:2014jba,JuarezPhDThesis}. In this same scenario, a recent study of 
two freely falling detectors showed that
both  entanglement and mutual information
can be acquired by the detectors  even when they  are causally disconnected by the event horizon \cite{Gallock-Yoshimura:2021yok}. An extension
of this scenario to a detector infalling toward a $(1+1)$-dimensional Cauchy horizon found that the transition rate of the infalling detector diverged
for a field in either the Unruh or Hartle-Hawking-Israel states~\cite{Juarez-Aubry:2021tae}.

Studies in $2+1$ dimensions have the advantage that one can work with exact solutions to the Einstein equations, unlike the aforementioned $(1+1)$-dimensional settings; furthermore, 
the detector can be coupled linearly to the field, instead of its momentum, without infrared complications. In spacetimes that are quotients of a maximally symmetric space, 
there is a further technical advantage insofar as the Wightman function of the quantum scalar field can be computed as a sum over images instead of the unavoidable mode sum in the Schwarzschild case \cite{Ng2022,Shallue:2025zto}. 
It was shown
some time ago~\cite{Hodgkinson:2012mr} that a UDW detector 
radially falling toward a static Ba\~nados-Teitelboim-Zanelli (BTZ) black hole exhibits the expected smooth and non-thermal response when the detector is still outside the horizon.  Recently, this study was extended to situations where the detector crosses the horizon~\cite{MariaRosaBTZ}. The response rate of the detector was found to be a smooth and slowly oscillating function, punctuated by non-differentiable points denoted as `glitches'.
Similar behaviour was found for the response rate of a detector falling into a rotating BTZ black hole
\cite{Wang:2024zny}. In this case four types of glitches were observed, with their locations dependent on the mass and angular momentum of the black hole. Taken together, these results
suggest that the event horizon of a black hole may be discernable to a local probe when quantum-field theoretic effects are included.

Here we investigate the effects of topology hidden behind the horizon on the transition rate of a freely falling detector before, at, and after the detector crosses the horizon. We consider the topological geon black hole spacetime known as the $\mathbb{R}\text{P}^{2}$ geon~\cite{Louko:1998hc}, which is a $\mathbb{Z}_2$ quotient 
of the spinless $(2+1)$-dimensional BTZ black hole, to which it is locally isometric, as well as being locally isometric to $(2+1)$-dimensional anti-de Sitter (AdS) spacetime.  

We begin in section~\ref{sec2} by describing the setup of the UDW detector in both the BTZ and geon spacetimes. For detector trajectories that are identical outside the horizon, we find that the Wightman functions for the scalar field differ on the two spacetimes due to the differing topologies behind the horizon. In sections \ref{analyticresults} and \ref{numericalresults} we compare the transition rates for the detector on the two spacetimes for a variety of initial conditions. 
If the detector's interaction is switched on in the exterior region, the transition rates on the BTZ black hole and the geon black hole remain qualitatively similar in the exterior, although the rate on the geon is larger; however, the rates differ considerably after the detector has crossed the black hole horizon. If the detector's interaction is switched on already when the detector is travelling up from the white hole, then qualitative differences appear already in the exterior region. 
In section~\ref{conc} we summarize our findings, 
concluding that an infalling detector is sensitive to hidden topology, albeit not in a dramatic manner.

\section{Setup}
\label{sec2}


\subsection{UDW detector model}

We consider a UDW detector \cite{Unruh_effect,DeWitt1979} modelled as a point-like two-level system,  
having states $|g\rangle$ and $|e\rangle$ with the respective energy eigenvalues $0$ and~$\Omega$. If $\Omega>0$, $\ket{g}$ is the ground state and $\ket{e}$ is the excited state; if $\Omega<0$, the roles of the two states are reversed. 

We couple the detector to a massless Klein-Gordon scalar field $\phi$ through the interaction Hamiltonian
\begin{equation}
    H_I = \lambda \chi(\tau) \mu(\tau) \otimes \phi\bigl(\mathsf{x}(\tau)\bigr),
\end{equation}
where $\lambda$ is the coupling constant, 
$\chi(\tau)$ is the switching function
that governs how the detector is switched on and off, $\mu = e^{i \Omega \tau} \sigma^+ + e^{-i \Omega \tau} \sigma^- $ is the detector's monopole moment operator with $\sigma^+ = |e\rangle \langle g|$ and $\sigma^- = |g\rangle \langle e|$, and $\mathsf{x}(\tau)$ is the trajectory of the detector parametrized by its proper time~$\tau$. 

Using first-order perturbation theory, with the initial state of the system being $|\psi_g\rangle = |g\rangle \otimes |0\rangle$ where $\ket{0}$ is the initial state of the field, 
the probability of the detector to make a transition to the state~$|e\rangle$, regardless of the final state of the field, is proportional to the response function, given by 
\begin{equation}\label{Fnodot}
    \mathcal{F}(\Omega) = \int \mathrm{d}\tau' \, \mathrm{d}\tau'' \, 
    \chi(\tau') \chi(\tau'') \, 
    \mathrm{e}^{-\ii \Omega(\tau' - \tau'')}W(\tau',\tau'') \,,
\end{equation}
where $W(\tau,\tau') := \bra{0} \phi\bigl(\mathsf{x}(\tau) \bigr)\phi\bigl(\mathsf{x}(\tau')\bigr)\ket{0}$ 
is the pullback to the detector's worldline of the field's Wightman function in the state $\ket{0}$ \cite{Kay:1988mu,Fewster:1999gj,Junker:2001gx,Louko:2007mu}. 
From now on we suppress the constant of proportionality and refer to the response function as the transition probability. 

Specialising to $2+1$ spacetime dimensions, we may take the switching function to be the characteristic function of an interval, 
\begin{equation}\label{chiex}
    \chi(t)= \begin{cases}1 & \tau_0 \leq t \leq \tau \,, \\ 0 & 
    \text{otherwise} \,, \end{cases}
\end{equation}
where $\tau_0$ and $\tau$ are the switch-on and switch-off moments, 
with $\tau_0 < \tau$. Differentiating $\mathcal{F}(\Omega)$ with respect to the switch-off moment $\tau$ gives then the detector's transition rate~\cite{Hodgkinson:2012mr}, 
\begin{equation}\label{Fdot}
\dot{\mathcal{F}}_\tau(\Omega) = \frac{1}{4} + 2 \int^{\Delta\tau}_{0}\,\mathrm{d}s \operatorname{Re}\left[ \mathrm{e}^{-\ii \Omega s}W(\tau,\tau-s)\right] \,,
\end{equation}
where $\Delta\tau := \tau-\tau_0$ denotes the total proper time that the detector operates. Despite the lack of smoothness in \eqref{chiex}, 
the transition rate is well defined when $\ket{0}$ is a Hadamard state, and it 
has a measurable interpretation in terms of an ensemble of identical detectors, all following the trajectory $\sx(\tau)$ and switched off at distinct moments 
\cite{Langlois:2005if,Louko:2007mu,Satz:2006kb}.

\subsection{BTZ spacetime}

The BTZ black hole spacetime is a vacuum solution of the $(2+1)$-dimensional Einstein equations with cosmological constant $\Lambda = -1/\ell^2$ where $\ell>0$ is the AdS length, and so the curvature of this spacetime is constant and negative everywhere \cite{BTZ1,BTZ2}. The metric in the exterior region of a spinless BTZ black hole has the form
\begin{equation}
ds^2 = -f(r) dt^2 + \frac{dr^2}{f(r)} + r^2 d\phi^2,
\label{BTZmetric}
\end{equation}
where $r>r_h = \ell \sqrt{M}$, 
$f(r)=( r^2/\ell^2 - M )$, $t \in (-\infty, \infty)$, 
$\phi \in [0,2\pi)$, and $M >0$ is the  (dimensionless) mass  of the black hole. The horizon is at $r=r_h$. The exterior is static, with the timelike hypersurface-orthogonal Killing vector~$\partial_t$. 

The metric \eqref{BTZmetric} can be obtained by 
starting from the AdS$_3$ spacetime, defined as the submanifold
\begin{equation}
    X_1^2 + X_2^2 -T_1^2-T_2^2 = -\ell^2
\end{equation}
embedded in the four-dimensional space $\mathbb{R}^{2,2}$ with metric
\begin{equation}
    ds^2 = dX_1^2+dX_2^2 - dT_1^2-dT_2^2 .
    \label{eq:metric_in_R_2_2}
\end{equation}
The region where $T_1 > |X_1|$ and $X_2 > |T_2|$ can be covered by new coordinates in which 
\begin{equation}
    \begin{aligned}
    X_1 & =\ell \frac{r}{r_h} \sinh \left(\frac{r_h}{\ell} \phi\right), & X_2 & =\ell \sqrt{\frac{r^2}{r_h^2}-1} \cosh \left(\frac{r_h}{\ell^2} t\right), \\
    T_1 & =\ell \frac{r}{r_h} \cosh \left(\frac{r_h}{\ell} \phi\right), & T_2 & =\ell \sqrt{\frac{r^2}{r_h^2}-1} \sinh \left(\frac{r_h}{\ell^2} t\right),
    \end{aligned}
    \label{AdstoBTZ}
\end{equation}
and this coordinate transformation brings the metric on AdS$_3$ to the form \eqref{BTZmetric} with $r \in\left(r_h, \infty\right)$, but with $\phi \in (-\infty,\infty)$.
This metric is adapted to a family of uniformly accelerated observers on AdS$_3$, and it may be described as the AdS$_3$-Rindler metric \cite{Jennings:2010vk,Henderson:2019uqo}. 

From this perspective, the 
exterior of the spinless 
BTZ spacetime can be obtained from AdS$_3$-Rindler spacetime by making an identification by the map
\begin{equation}
    \Gamma: (t, r, \phi) \mapsto (t, r, \phi+2 \pi) \,, 
\label{Gamma-id}
\end{equation}
so that $\phi$ becomes an angular coordinate. 
The perspective further extends to the 
region of AdS$_3$ where $T_1 > |X_1|$: 
this region can be covered by the coordinates $(U,V,\phi)$, where $-1 < UV<1$ and $\phi\in\mathbb{R}$, by 
\begin{align}
X_1 & =\ell\left(\frac{1-U V}{1+U V}\right) \sinh \left(\frac{r_h}{\ell} \phi\right), & X_2 & =\ell \frac{V-U}{1+U V}, \notag \\
T_1 & =\ell\left(\frac{1-U V}{1+U V}\right) \cosh \left(\frac{r_h}{\ell} \phi\right), & T_2 & =\ell \frac{V+U}{1+U V},
\end{align}
in which the metric reads 
\begin{equation}
d s^2=-\frac{4 \ell^2}{(1+U V)^2} d U d V
+r_h^2\left(\frac{1-U V}{1+U V}\right)^2 d \phi^2,
\label{eq:kruskalchart}
\end{equation}
the extension of \eqref{Gamma-id} reads 
\begin{equation}
    \Gamma: (U, V, \phi) \mapsto (U, V, \phi+2 \pi)  \,,
\label{Gammaprime-id}
\end{equation}
and the identification by \eqref{Gammaprime-id}
gives the extended spinless BTZ black-and-white hole spacetime. 
The Penrose-Carter diagram is shown in Figure~\ref{fig:Penrose_btz}.
The exterior \eqref{BTZmetric} is the region where $U<0$ and $V>0$, wherein the coordinate transformation from $(U,V,\phi)$ to $(t,r,\phi)$ reads
\begin{equation}
\frac{r}{r_h}=\frac{1-U V}{1+U V} \ , \quad \frac{r_h t}{\ell^2}=\ln \sqrt{-\frac{V}{U}} \,.
\end{equation}
The exterior timelike Killing vector $\partial_t$ extends to the whole BTZ black hole spacetime as $r_h \ell^{-2} (V\partial_V - U \partial_U)$, which is timelike in the two exteriors, spacelike in the black hole and white hole interiors, and null on the Killing horizons separating these four quadrants.

\subsection{Geon spacetime}

The $\mathbb{R}\text{P}^{2}$ geon black hole (or more simply geon) is obtained from the spinless BTZ spacetime by an identification under the map 
\begin{equation}
J: (U, V, \phi) \mapsto (V, U, \phi+\pi).
\label{eq:geon-Jmap}
\end{equation} 
Since $\phi \in [0,2\pi)$ on the BTZ spacetime, it is evident that $J^2 = Id_{\text{BTZ}}$, and therefore $J$ generates a $\mathbb{Z}_2 \simeq\{Id_{\text{BTZ}}, J\}$ action on the BTZ spacetime. 
Therefore, the geon spacetime is the quotient spacetime
$\mathcal{M}_{\text {geon }}=\mathcal{M}_{\text{BTZ}} / \mathbb{Z}_2$. 

\begin{figure}[t]
    \centering
    \includegraphics[width=0.8\linewidth]{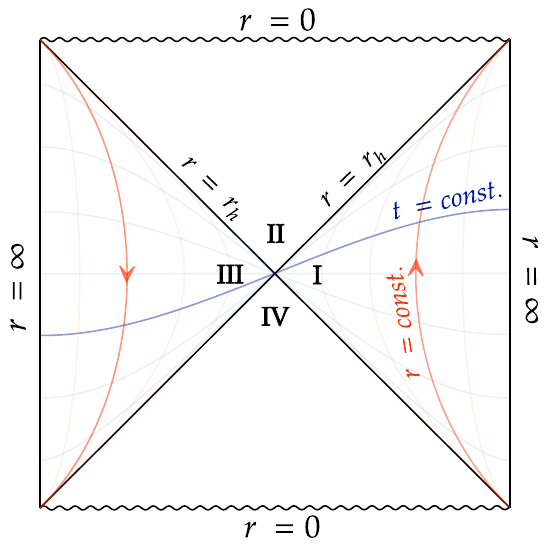}
    \caption{Penrose-Carter diagram of the spinless BTZ black hole~\cite{BTZ2}. 
    The suppressed dimension is the $2\pi$-periodic coordinate~$\phi$. 
    Region I is an exterior, covered by the metric~\eqref{BTZmetric}, and curves of constant $r$ and constant $t$ are shown. Region III is a second exterior, covered by a similar coordinate chart, and the extension of the surfaces of constant $t$ therein are shown. Region II is the black hole interior and Region IV is the white hole interior. The four quadrants are separated by the horizon, where $r=r_h$. The arrows represent the direction of the globally-defined Killing vector 
    $r_h \ell^{-2} (V\partial_V - U \partial_U)$, 
    which is timelike in regions I and III and spacelike in regions II and~IV\null. In~I, this Killing vector is given by~$\partial_t$.}
    \label{fig:Penrose_btz}
\end{figure}

\begin{figure}[t]
    \centering
    \includegraphics[width=0.5\linewidth]{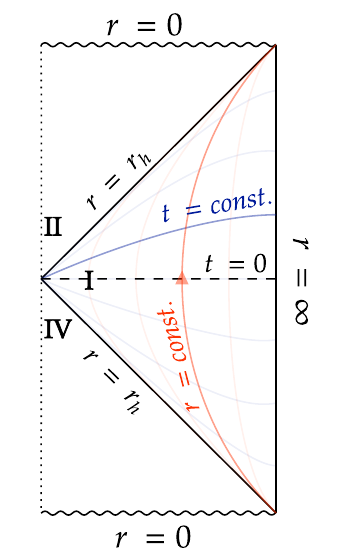}
    \caption{Penrose-Carter diagram of the geon spacetime. The diagram consists of the right half of that in Figure~\ref{fig:Penrose_btz}. The suppressed dimension is again the coordinate~$\phi$, which has period $2\pi$ everywhere except at the vertical dotted line on the left, where $\phi$ has period~$\pi$. Region I is isometric to region I of the BTZ hole, and it has the timelike Killing vector~$\partial_t$, but this Killing vector does not extend to a globally-defined Killing vector on the full geon spacetime. 
    The global geometry of the geon hence gives region I a distinguished constant $t$ hypersurface, shown in the figure as the hypersurface $t=0$.}
    \label{fig:Penrose_geon}
\end{figure}

In the Penrose-Carter diagram of the BTZ spacetime, shown in Figure~\ref{fig:Penrose_btz}, $J$ \eqref{eq:geon-Jmap} acts by a left-right reflection in the dimensions shown, composed with the antipodal map in the suppressed circle. It follows that the Penrose-Carter diagram of the geon consists of the (say) right half of the BTZ diagram, as shown in Figure~\ref{fig:Penrose_geon}. The geon is time orientable, and it has the nonorientable spatial topology $\mathbb{R}\text{P}^{2} \setminus\{\text{point at infinity}\}$. 

As seen from Figure~\ref{fig:Penrose_geon}, 
the geon is an eternal black-and-white hole spacetime. It has a single exterior region, isometric to an exterior of the BTZ hole, and this region is static, with the timelike hypersurface-orthogonal Killing vector~$\partial_t$. Because of the unusual topology behind the horizons, however, $\partial_t$ does not extend to a globally-defined Killing vector on the whole geon spacetime: in Figure~\ref{fig:Penrose_geon}, the extension has a sign inconsistency at the vertical dotted line on the left. The topology behind the horizons therefore endows the exterior with a distinguished hypersurface of constant~$t$, shown in Figure \ref{fig:Penrose_geon} as $t=0$: this is the only constant $t$ hypersurface that extends from the exterior smoothly to the horizon. The distinguished hypersurface cannot be identified by active classical observations in the exterior, and this exemplifies the topological censorship theorem~\cite{Friedman_1995,Galloway:1996hx,Galloway:1999bp}. 
The hypersurface can however be identified by quantum observations of a state that has been adapted to the geon's global geometry, to which we now turn. 

\subsection{The Wightman function for the AdS, BTZ and geon spacetimes}

We consider a detector interacting with a conformally coupled massless scalar field $\phi(x)$, that is, the field satisfies the Klein-Gordon equation $(\Box - R/8)\phi(x) = 0$, where $\Box$ is the d'Alembert operator and $R = -6/\ell^2$ is the 
Ricci scalar~\cite{Wald:1984rg}. The numerical coefficient $1/8$ is the conformal coupling value $\frac14 (d-2)/(d-1)$ in $d=3$ spacetime dimensions. 

In order to study the transition rate of a UDW detector, 
the Wightman function in the field's initial state must be computed. 
For both the BTZ spacetime and the geon spacetime, we consider the initial state that is induced by the global vacuum on AdS${}_3$. As both the BTZ spacetime and the geon spacetime are quotients of an open region on AdS${}_3$ by a discrete isometry group, the strategy is to use the method of images. 

Recall that the Wightman function in the global vacuum on AdS${}_3$ reads \cite{Lifschytz_1994} 
\begin{equation}
W_{\text{AdS}_3}\left(x, x^{\prime}\right)=\frac{1}{4 \pi \sqrt{2} \ell}\left(\frac{1}{\sqrt{\sigma\left(x, x^{\prime}\right)}}-\frac{\zeta}{\sqrt{\sigma\left(x, x^{\prime}\right)+2}}\right) , 
\label{eq:AdS_Wightman}
\end{equation}
where 
\begin{align}
    \sigma(x, x^{\prime}) &= \frac{1}{2 \ell^2}\left[\left(X_1-X_1^{\prime}\right)^2-\left(T_1-T_1^{\prime}\right)^2 \right. 
    \notag \\
    & \hspace{3ex} \left. +\left(X_2-X_2^{\prime}\right)^2-\left(T_2-T_2^{\prime}\right)^2\right] , 
    \label{eq:geodesic_distance}
\end{align}
and the parameter $\zeta \in \{1,-1,0\}$ specifies the boundary condition at the asymptotically AdS infinity. 
$\zeta =1$ is the Dirichlet condition and $\zeta = -1$ is the Neumann condition, and both of these define a unitary quantum theory, with no probability entering or exiting through the infinity. 
$\zeta=0$ is known as the transparent boundary condition, which classically allows probability to exit the spacetime through the infinity, and has a less clear physical interpretation in the quantum theory. 

Note that $\sigma(x, x^{\prime})$ \eqref{eq:geodesic_distance} is half of the dimensionless squared geodesic distance in the embedding space $\mathbb{R}^{2,2}$ \eqref{eq:metric_in_R_2_2} \cite{Hodgkinson:2012mr}; 
we shall henceforth refer to 
$\sigma$ simply as the \textit{geodesic distance}.
Note also that $W_{\text{AdS}_3}$ has distributional singularities at the zeroes of the denominators in~\eqref{eq:AdS_Wightman}, 
and its full definition includes an $i\epsilon$ prescription for these singularities and the branch points of the square roots. We shall address this shortly. 

Now, the Wightman functions that $W_{\text{AdS}_3}$ induces on the BTZ spacetime and on the geon spacetime are, respectively \cite{Lifschytz_1994,carlip19952+}, 
\begin{subequations}
\label{WBTZ-and-geon}
\begin{align}
W_{\text{BTZ}}\left(x, x^{\prime}\right)  &= 
\sum_{n=-\infty}^{\infty} W_{\text{AdS}_3}\left(x, \Gamma^n x^{\prime}\right) , 
\label{WBTZ}
\\
W_{\text{geon}}\left(x, x^{\prime}\right) &= 
W_{\text{BTZ}}\left(x, x^{\prime}\right)
+W_{\text{BTZ}}\left(x, J x^{\prime}\right) . 
\label{Wgeon}
\end{align}
\end{subequations}
$W_{\text{BTZ}}$ has the characteristics of a Hartle-Hawking-Israel (HHI) vacuum \cite{Hartle:1976tp,Israel:1976ur}, 
describing a black hole in thermal equilibrium with a heat bath at infinity~\cite{Lifschytz_1994}. 
In particular, the restriction of $W_{\text{BTZ}}$ to an exterior region of the BTZ spacetime is stationary, because the map $\Gamma$ \eqref{Gamma-id} on the AdS${}_3$-Rindler spacetime commutes with the Killing vector 
$r_h \ell^{-2} (V\partial_V - U \partial_U)$ that generates the exterior Killing time translations. 
By contrast, $W_{\text{geon}}$ does not describe a thermal equilibrium state on the geon. The second term in \eqref{Wgeon} shows that $W_{\text{geon}}$ differs from $W_{\text{BTZ}}$ already in the geon's exterior, and this term further shows that the restriction of $W_{\text{geon}}$ to the geon's exterior is not invariant under the exterior Killing time translations, since the quotienting map $J$ \eqref{eq:geon-Jmap} does not commute with the BTZ Killing time translations. It hence follows that $W_{\text{geon}}$ cannot describe a thermal equilibrium state, although it still has some characteristics of a HHI state, as analysed in \cite{Louko:1998dj,Louko:1998hc,Louko:2000tp,Guica:2014dfa,Smith_2014}. In particular, the response of a static detector in the geon's exterior is not stationary, in contrast to the response of a static detector in the BTZ exterior, although the two asymptotically agree for a detector that operates at times much earlier than the $t=0$ hypersurface shown in Figure \ref{fig:Penrose_geon} \cite{Smith_2014}. 

We shall proceed to investigate how the responses differ for an infalling detector. As preparation, we record here that when the points $x$ and $x'$ are in the exterior region, equations \eqref{AdstoBTZ}, \eqref{Gamma-id}, \eqref{eq:geon-Jmap} and \eqref{eq:geodesic_distance} give 
\begin{widetext}
\begin{subequations}
\label{sigma_BTZandgeon}
\begin{align}
    \sigma(x, \Gamma^n x^{\prime}) &= \frac{r r^{\prime}}{r_h^2} \cosh \! \left(\frac{r_h}{\ell}(\Delta \phi-2 \pi n)\right)-1  -\frac{\left(r^2-r_h^2\right)^{\frac{1}{2}}\left(r^{\prime 2}-r_h^2\right)^{\frac{1}{2}}}{r_h^2} \cosh \! \left(\frac{r_h}{\ell^2} \Delta t -i\epsilon \right) , 
    \label{sigma_BTZ}
\\
    \sigma(x, J \Gamma^n x^{\prime}) &= \frac{r r^{\prime}}{r_h^2} \cosh \! \left(\frac{r_h}{\ell}\left(\Delta \phi-2 \pi n 
    - \pi \right)\right)-1  +\frac{\left(r^2-r_h^2\right)^{\frac{1}{2}}\left(r^{\prime 2}-r_h^2\right)^{\frac{1}{2}}}{r_h^2} 
    \cosh \! \left(\frac{r_h}{\ell^2} (t+t^{\prime})\right) , 
    \label{sigma_geon}
\end{align}
\end{subequations}
\end{widetext}
where $\Delta \phi = \phi - \phi^\prime$ and
$\Delta t = t-t^\prime$, and the $\epsilon\to0^+$ limit in \eqref{sigma_BTZ} specifies the distributional character of the Wightman function. 
Note that \eqref{sigma_BTZ} depends on $t$ and $t'$ only via the combination~$\Delta t$, whereas \eqref{sigma_geon} depends on $t$ and $t'$ only via the combination $t+t'$. This shows explicitly that $W_{\text{BTZ}}$ is invariant under the exterior Killing time translations but $W_{\text{geon}}$ is not.


\subsection{Free-falling detector}

Our interest  is in computing, in different parameter regimes, the transition rate of a UDW detector that is radially free-falling into the geon and comparing it to its BTZ counterpart. We are particularly interested in the transition rate of the detector near and beyond the horizon.

In this scenario, the detector's trajectory in the exterior region is given by 
\cite{Hodgkinson:2012mr}
\begin{equation}
\begin{aligned}
& r=\ell \sqrt{M} q \cos \tilde{\tau},
\\
& t=(\ell / \sqrt{M}) \operatorname{arctanh} \! \left(\frac{\tan \tilde{\tau}}{\sqrt{q^2-1}}\right) + t_0 ,\\
&\phi=\phi_0,
\end{aligned}
\label{trajectory}
\end{equation}
where $\tilde{\tau} := \tau/\ell$ is a dimensionless affine parameter, 
and $q>1$, $t_0\in \mathbb{R}$ and $\phi_0 \in [0, 2\pi)$ are constants. 
$t_0$~is the value of the exterior Killing time at which $r$ reaches its maximum value. 
$\phi_0$ is the constant value of $\phi$ on the trajectory. 

Inserting \eqref{trajectory} into \eqref{sigma_BTZ} with $\sigma_n^{\text{BTZ}}\left(\tau,\tau^{\prime}\right) := \sigma\left(x_D(\tau), \Gamma^n x^{\prime}_D(\tau^{\prime})\right)$ and into \eqref{sigma_geon} with $\sigma_n^{\text{geon}}\left(\tau,\tau^{\prime}\right) := \sigma\left(x_D^\mu(\tau), J \Gamma^n x_D^\mu\left(\tau^{\prime}\right)\right)$, we obtain
\begin{subequations}
\begin{align}
\sigma_n^{\text{BTZ}}\left(\tau,\tau^{\prime}\right) 
&=-1 + K^{\text{BTZ}}_n \cos(\tilde{\tau})\cos(\tilde{\tau}^{\prime}) + \sin(\tilde{\tau})\sin(\tilde{\tau}^{\prime}) , 
\label{sigma_BTZ_fall}
\\
\sigma_n^{\text{geon}}\left(\tau,\tau^{\prime}\right)
&=-1+A^n_{t_0}(\tau) \cos \left(\tilde{\tau}^{\prime}\right)+ {B_{t_0}(\tau) \sin \left(\tilde{\tau}^{\prime}\right)},
\label{sigma_geon_fall}
\end{align}
\end{subequations}
where
\begin{subequations}
\begin{align}
A^n_{t_0}(\tau) & = K_{n,t_0}^{\text{geon}} \cos (\tilde{\tau})
+\sinh \!\left(2 \sqrt{M} \tilde t_0\right) \sqrt{q^2-1} \sin (\tilde{\tau}) , \\
B_{t_0}(\tau) & = \cosh \! \left(2 \sqrt{M} \tilde t_0\right) \sin (\tilde{\tau})
\notag 
\\
&\hspace{3ex} 
+\sinh \! \left(2 \sqrt{M}\tilde t_0\right) \sqrt{q^2-1} \cos (\tilde{\tau}),
    \label{AnBn}
\end{align}
\end{subequations}
and 
\begin{subequations}
\begin{align}
K_n^{\text{BTZ}} & = 1 + q^2 \sinh^2 \! \left(n \pi \sqrt{M}\right) , 
\label{Knbtz}
\\
K_{n,t_0}^{\text{geon}} & = q^2 \cosh \! \left((2n+1) \pi \sqrt{M}\right)
\notag 
\\
&\hspace{3ex}+\cosh \! \left(2 \sqrt{M} \tilde t_0\right) (q^2-1) , 
\label{Kngeon}
\end{align}
\end{subequations}
with $\tilde{t}_0 = t_0/\ell$. 
The constant $\phi_0$ does not appear, as must be the case by the invariance of the spacetimes and the quantum states under rotations in~$\phi$. 
The constant $t_0$ has disappeared from the expressions relevant for the BTZ spacetime, as must be by the invariance under the Killing time translations, but it remains in expressions relevant for the geon spacetime, which reflects the absence of global Killing time translations on the geon. 

Using the above expressions, we find that the transition rate on the BTZ spacetime is given by 
\begin{widetext}
\begin{multline}
\dot{\mathcal{F}}_\tau^{\text{BTZ}}(\Omega) = \frac14
        +\frac{1}{2 \pi \sqrt{2}} \sum_{n=-\infty}^{\infty} \int_0^{\Delta \tilde{\tau}} \mathrm{d} \tilde{s} \operatorname{Re}
        \left[\frac{\mathrm{e}^{-\ii \tilde{\Omega} \tilde{s}}}{\sqrt{-1+K_n^{\text{BTZ}} \cos \tilde{\tau} \cos (\tilde{\tau}-\tilde{s})+\sin \tilde{\tau} \sin (\tilde{\tau}-\tilde{s})}}\right. \\
        \left.-\zeta \frac{\mathrm{e}^{-\ii \tilde{\Omega} \tilde{s}}}{\sqrt{1+K_n^{\text{BTZ}} \cos \tilde{\tau} \cos (\tilde{\tau}-\tilde{s})+\sin \tilde{\tau} \sin (\tilde{\tau}-\tilde{s})}}\right] , 
        \label{btz-transition_rate}
\end{multline}
and that on the geon is given by 
\begin{align}
\dot{\mathcal{F}}_\tau^{\text{geon}}(\Omega)= \dot{\mathcal{F}}_\tau^{\text{BTZ}}(\Omega)+ \Delta\dot{\mathcal{F}}_\tau(\Omega) , 
\end{align}
where 
\begin{multline}
\Delta\dot{\mathcal{F}}_\tau(\Omega)=
         \frac{1}{2 \pi \sqrt{2}}  \sum_{n=-\infty}^{\infty} \int_0^{\Delta \tilde{\tau}} \mathrm{d} \tilde{s} \operatorname{Re}\left[\frac{\mathrm{e}^{-i \tilde{\Omega} \tilde{s}}}{\sqrt{-1+A^n_{t_0}(\tau) \cos \left(\tilde{\tau}-\tilde{s}\right)+ {B_{t_0}(\tau) \sin \left(\tilde{\tau}-\tilde{s}\right)}}}\right. \\
        \left.-\zeta \frac{\mathrm{e}^{-i \tilde{\Omega} \tilde{s}}}{\sqrt{1+A^n_{t_0}(\tau) \cos \left(\tilde{\tau}-\tilde{s}\right)+ {B_{t_0}(\tau) \sin \left(\tilde{\tau}-\tilde{s}\right)}}}\right],   
        \label{geon-transition_rate}
\end{multline}
\end{widetext}
where $\tilde\Omega = \ell\Omega$, $\Delta \tilde{\tau} = (\tau-\tau_0)/\ell$, and $\tau_0$ is the proper time at which the detector is switched on.
The square roots in \eqref{btz-transition_rate} and \eqref{geon-transition_rate} are positive for positive arguments, and they are analytically continued to negative arguments  by giving $\tilde{s}$ a small negative imaginary part, as follows from the distributional properties of the Wightman function~\cite{Hodgkinson:2012mr,Kay:1988mu}. 

Although equations \eqref{sigma_BTZandgeon} are only valid exterior to the black hole, the transition rate formulas \eqref{btz-transition_rate}--\eqref{geon-transition_rate} hold also for detectors that operate even before exiting the white hole and/or after entering the black hole. This follows by analytic continuation, since both the BTZ hole and the geon have global analytic charts, as seen from~\eqref{eq:kruskalchart}.


\section{Analytic results}
\label{analyticresults}

While both 
$\dot{\mathcal{F}}_\tau^{\text{BTZ}}$
and 
$\dot{\mathcal{F}}_\tau^{\text{geon}}$ 
are well defined, their derivatives with respect to $\tau$ have discontinuities at values of $\tau$ where a denominator in the integrand in 
\eqref{btz-transition_rate} or \eqref{geon-transition_rate} 
is singular at an endpoint of integration. 
We refer to these discontinuities in the derivative as~`glitches'~\cite{MariaRosaBTZ}. 
In this section we present analytic estimates of where these glitches occur for the BTZ spacetime and for the geon spacetime.

\subsection{BTZ glitches}

Consider first the geodesic distance in BTZ spacetime. 

For $n=0$, $K_n^{\text{BTZ}}=K_0^{\text{BTZ}} = 1$, and $\sigma^{\text{BTZ}}_0(\tau,\tau')$ only equals zero if the points at $\tau$ and $\tau'$ are null-separated. This occurs  in \eqref{btz-transition_rate} at the lower limit of integration  ($\tilde{s}$=0), where the two points in fact coincide, and it does not introduce any singularities nor any need for analytic continuation. 

The situation is different for $n\neq 0$ because, physically, this represents the distance between a point on the trajectory and a point on an identical trajectory but translated by $\phi \xrightarrow{} \phi + 2 \pi n$ in AdS spacetime. Therefore, by choosing $\tau \text{ and } \tau'$ appropriately, it is always possible to find two points on the two trajectories that are null-separated if they are in causally connected regions of spacetime.

The argument of the square root in the $\zeta$-dependent term is positive for all $\tilde{s}$ in the interval $0<\tilde{s}\le\Delta\tilde{\tau}<\pi/2$ and all $n\ge1$. 
However in the first term, the argument of the square root can change sign within the interval of integration for $n\ge1$.  
This results in a singularity in the integrand, but the singularity is of the inverse square root type and hence integrable.
Analytically solving the equation $\sigma^{\text{BTZ}}_n(\tau,\tau-s)=0$, 
we obtain
\begin{widetext}
\begin{equation}
s_{\text{BTZ}}\left(n, \tau\right):= \tau- \ell \arctan \! \left(\frac{\tan(\tilde\tau)-\cos(\tilde\tau) K^{\text{BTZ}}_n \sqrt{(K^{\text{BTZ}}_n)^2-1}}{K^{\text{BTZ}}_n+\sin(\tilde\tau) \sqrt{(K^{\text{BTZ}}_n)^2-1}}\right).
\end{equation}

Examining $s_{\text{BTZ}}\left(n, \tau\right)$, we find that for $\tilde\tau \in [0,\pi/2]$,  $s_{\text{BTZ}}\left(n, \tau\right) \in  [0,\pi\ell]$ for all $n \in \mathbb{N}$. 
Since within the first integral in (\ref{btz-transition_rate}) the integration variable $\tilde{s} \in [0,\Delta \tilde{\tau}]$, there will be a time $\tau^{\text{BTZ}}_n$  at which the singularity in the integrand coincides with the endpoint of the integration interval $\Delta\tau$, given by the expression
\begin{equation}
\tau^{\text{BTZ}}_n - \tau_0= s_{\text{BTZ}}\left(n, \tau^{\text{BTZ}}_n \right) 
\end{equation}
which is equivalent to directly solving the equation:
\begin{equation}
\sigma^{\text{BTZ}}_n(\tau^{BTZ}_n,\tau_0)=0.
\label{glitches_definition}
\end{equation} 
Solving this analytically yields 
\begin{equation}
\tau^{\text{BTZ}}_n= 
\ell \text{sign}\left(\tilde\tau_0 + \arctan\sqrt{(K^{\text{BTZ}}_n)^2 - 1}\right)
\arccos\left(\frac{K^{\text{BTZ}}_n\sqrt{1+ \tan^2(\tilde\tau_0)}-\sqrt{(K^{\text{BTZ}}_n)^2-1} \tan(\tilde\tau_0)}{(K^{\text{BTZ}}_n)^2+\tan^2(\tilde\tau_0)}\right) .
\label{btzglitch_tau0}
\end{equation}
\end{widetext}
These points $\tau^{\text{BTZ}}_n$ are visible in the transition rate as points of non-differentiability; these are the 
`glitches' noted above. However the transition rate remains well defined at these points.

In the case where $\tau_0 = 0$, 
in which the detector is turned on at the moment when $r$ reaches its maximum value,  
the expression \eqref{btzglitch_tau0} simplifies to \cite{MariaRosaBTZ} 
\begin{equation}
\tau^{\text{BTZ}}_n= \ell\arccos\left(\frac{1}{K^{\text{BTZ}}_n}\right).
\label{btzglitch}
\end{equation}

It is interesting to note that these  glitches can be detected even before the detector crosses the event horizon.   This follows from the fact that, if the parameters are such that   $K^{\text{BTZ}}_n < q$, then  $\tau^{BTZ}_n < \tau_H$, where $\tau_H = \ell\arccos(1/q)$ is the 
instant at which the detector crosses the event horizon.

It is worth noting that the transition rate calculated with the $n=0$ term alone has a simplified expression, found in Eq. (6.5) of \cite{{Hodgkinson:2012mr}}. Also, the $n\ne0$ terms do not change under $n\to-n$, so the transition rate can be written as twice the sum over $n\ge1$.

\subsection{Geon glitches}

The transition rate associated with the geon black hole consists of two terms. One term is identical to the BTZ rate (\ref{btz-transition_rate}); the other is
given by \eqref{geon-transition_rate}.
This extra term is obtained 
from the BTZ rate (\ref{btz-transition_rate}) by substitution of (\ref{sigma_geon_fall}) in place of (\ref{sigma_BTZ_fall}). 
This means that the glitches present in the transition rate of a detector  in the geon spacetime will be the same as those in the BTZ case, plus additional glitches specific to the geon, obtained by solving  $\sigma^{\text{geon}}_n(\tau^{\text{geon}}_n,\tau_0)=0$,
or, for $\zeta=\pm1$, also $\sigma^{\text{geon}}_n(\tau^{\text{geon}}_n,\tau_0)+2=0$. Geometrically, the second case corresponds to a null geodesic that connects the switch-on event to a later event on the trajectory after reflection from  infinity. In this section we consider only the first of these cases, but we shall return to the second case in Section~\ref{subsec:zetanonzero}. 
We note from \eqref{Kngeon} that the sum in \eqref{geon-transition_rate} is invariant under $n\to -1-n$, so it suffices to sum over $n\ge0$ and double the result.

\begin{figure}[!p]
    \centering
    \includegraphics[width=0.75\linewidth]{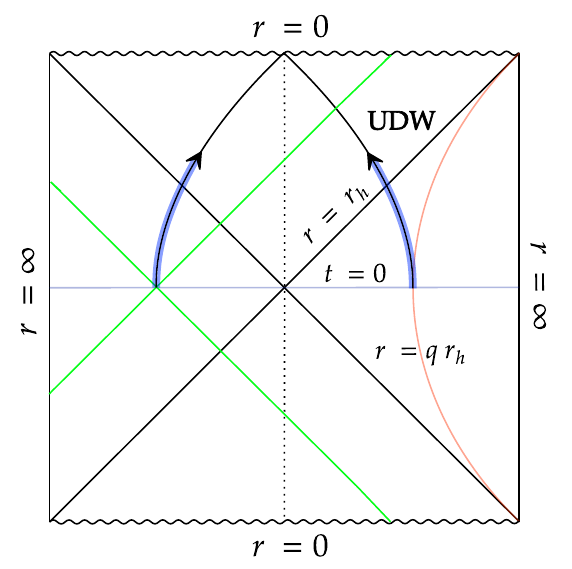} 
    \includegraphics[width=0.75\linewidth]{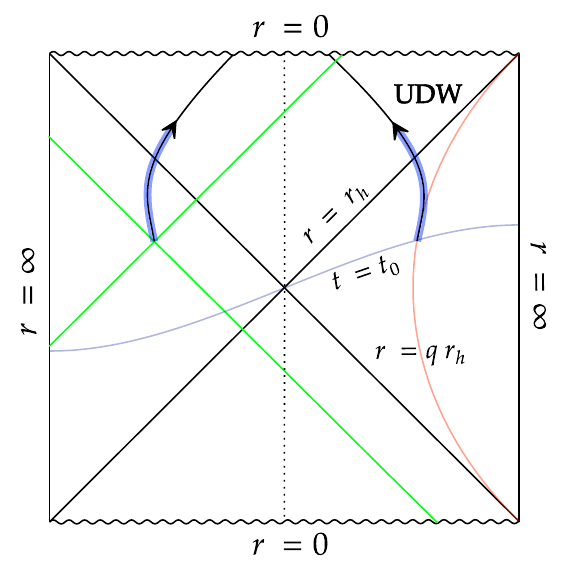}
   \includegraphics[width=0.75\linewidth]{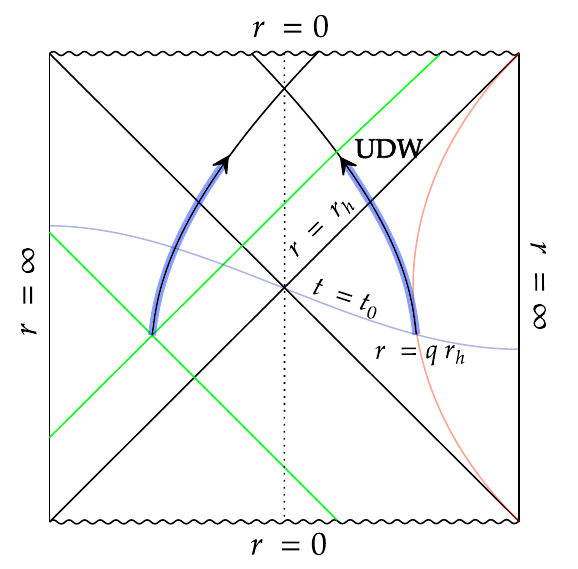}
    \caption{Causal relationship between points on the infalling worldline on the geon spacetime, as depicted on the BTZ spacetime by the trajectory on the right and its image under the map $J$ \eqref{eq:geon-Jmap} on the left. All three panels have $\tau_0=0$, so that the detector is switched on at the maximum value of $r$ on the trajectory, but this maximum value occurs at $t_0=0$ (upper panel), $t_0>0$ (middle panel) and $t_0<0$ (lower panel). The green lines represent light cones of the point on the image trajectory at which the detector is switched on.}
    \label{fig:Penrose_analysis_t0}
\end{figure}

\begin{figure}[!t]
    \centering
    \includegraphics[width=0.75\linewidth]{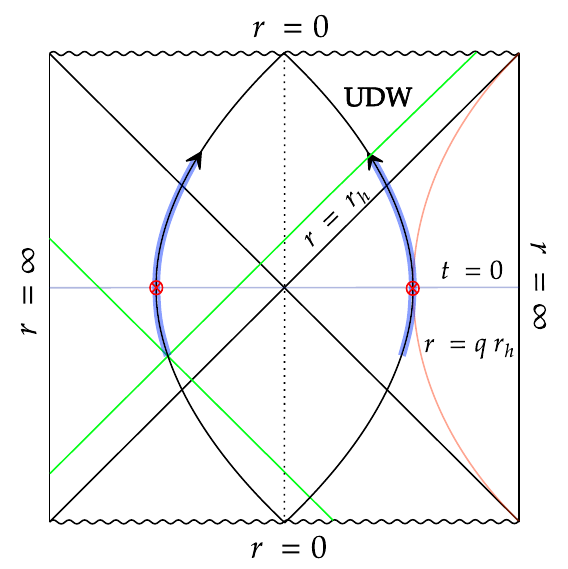}
    \includegraphics[width=0.75\linewidth]{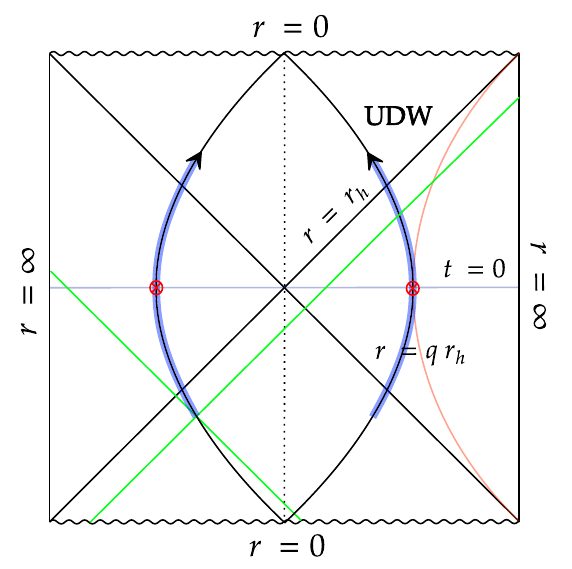}   
    \caption{As in Figure \ref{fig:Penrose_analysis_t0} but with $t_0=0$ and $\tau_0<0$, the upper panel with $-\tau_H < \tau_0 < 0$ and the lower panel with $\tau_0 < -\tau_H$. Note that in the lower panel the detector is switched on already in the white hole interior. The red point represents the position of the detector at the proper time $\tau=0$.} 
\label{fig:Penrose_analysis_tau0}
\end{figure}

Where are these latter glitches located? In general, there are different cases depending on the value of the parameters $t_0$ and $\tau_0$. This can be understood by analyzing the Penrose-Carter diagram of the BTZ black hole; in this spacetime, (\ref{sigma_geon_fall}) represents the distance between a point on the detector's trajectory and a point on its mirror image  with a shift in $\phi$ by $(2n+1)\pi$. 
Geon glitches  occur when a null ray from the image detector emitted at proper time $\tau_0$ intersects the trajectory of the original detector in the geon covering space.
We can qualitatively predict what happens to the geon glitches for different values of $t_0$ and $\tau_0$, keeping in mind that in the Penrose-Carter diagram the $\phi$ coordinate is suppressed, so what we can draw are the null rays that travel radially. In this analysis the intersection between the null ray and the detector's trajectory does not give us the exact position of the glitches, but does give a bound as to where the geon glitches can or cannot occur; the actual null rays have to travel through the $\phi$ direction by an odd multiple of~$\pi$.  
This approach gives us a good protocol for gaining insight into the transition rate without relying on analytical expressions for the glitches. Additionally, it serves as a consistency check for our numerical results.

We consider in turn the case where $t_0 \neq 0 = \tau_0$, the case where $t_0 = 0 \ne \tau_0$, and the case where $t_0 = 0 = \tau_0$. The features of the generic case, in which $t_0 \neq 0 \ne \tau_0$, can be inferred from these.

\subsubsection{$t_0 \neq 0 = \tau_0$}

Consider first the case in which $t_0 \neq 0 = \tau_0$. 
As illustrated in Figure~\ref{fig:Penrose_analysis_t0}, changing the value of $t_0$ generates the three distinct scenarios described below.

\begin{itemize}

    \item $t_0 = 0$: From the  top diagram in Figure~\ref{fig:Penrose_analysis_t0}, we observe that a null ray from the initial location of the image detector (at the left) encounters the trajectory of the original detector (at the right), so it is possible   to detect glitches in this regime, albeit only inside the horizon as shown in the diagram.

    \item $t_0 > 0$: From the middle diagram in Figure~\ref{fig:Penrose_analysis_t0}, we  {observe} that geon glitches can be observed from within the horizon. However 
    the first geon glitch  occurs closer to the future singularity than in the $t_0=0$ case. 
    By continuity there  is a time  $t_0$ such that no geon glitches will be observable before the detector encounters the singularity.
    
    \item $t_0 < 0$: From the bottom diagram in Figure~\ref{fig:Penrose_analysis_t0}, we see that null rays from the image detector at the left can reach the original infalling detector before either encounters the singularity. Multiple geon glitches can therefore occur, but all of them will be behind the event horizon.
    Moreover, we expect  that $ \lim_{t_0\to-\infty} \tau^{\text{geon}}_0 = \tau_H$.
\end{itemize}

It is easy to find the position of the geon glitches in analytic form.  Following the same definition adopted for the BTZ case, we obtain
\begin{widetext}
\begin{equation}
\tau^{\text{geon}}_n(t_0)= \ell \arccos\left(\frac{K_{n,t_0}^{\text{geon}}-\sqrt{q^2-1} \sqrt{\Delta^2-1} \sinh{\left(2 \sqrt{M} \tilde t_0\right)} }{\Delta^2}\right) , 
\label{geonglitch_t0}
\end{equation}
\end{widetext}
with $\Delta^2 = (K_{n,t_0}^{\text{geon}})^2+\left(q^2-1\right) \sinh^2 \! \left(2 \sqrt{M} \tilde t_0\right)$.
It can be easily seen that 
\begin{subequations}
\begin{align}
\lim_{t_0 \rightarrow -\infty} \tau^{\text{geon}}_n &= \tau_H , 
\\
\lim_{t_0 \rightarrow 0} \tau^{\text{geon}}_n &= \ell \arccos \! \left(\frac{1}{K_{n,0}^{\text{geon}}}\right).
\end{align} 
\end{subequations}
By analyzing \eqref{geonglitch_t0} it is evident that there are also solutions for $t_0>0$ and that the position of the glitches grows with~$n$. 
As $t_0$ increases, eventually the argument of the arccos in  \eqref{geonglitch_t0} is equal to zero,
at $t=t_0^{lim}$; 
for larger values of $t_0$  the glitches reach the singularity before they can be detected along the trajectory. As we can infer from the middle diagram in Figure \ref{fig:Penrose_analysis_t0}, as $t_0$ increases, the location at which the detector encounters the singularity moves increasingly rightward, and 
 no more glitches can occur for $t_0>t_0^{lim}$, where 
\begin{equation}
    t_0^{lim}= \frac{\ell}{2\sqrt{M}} \arccosh \! \left({\frac{q}{\sqrt{q^2-1}}}\right) .
    \label{tlim}
\end{equation}

\subsubsection{$t_0 = 0 \ne \tau_0$}

Consider next the case in which $t_0 = 0 \ne \tau_0$. 
The geodesic distance (\ref{sigma_geon_fall}) becomes 
\begin{equation}
\begin{aligned}         
\sigma_n^{\text{geon}}\left(\tau,\tau^{\prime}\right)=
    -1 + K^{\text{geon}}_n \cos(\tilde{\tau})\cos(\tilde{\tau}^{\prime}) + \sin(\tilde{\tau})\sin(\tilde{\tau}^{\prime}),
\end{aligned}
\label{sigma_geon_std}
\end{equation}
with $K^{\text{geon}}_n := K^{\text{geon}}_{n,0} $.
This expression is identical to~\eqref{sigma_BTZ_fall}, with the substitution of the function \eqref{Kngeon} in place of~\eqref{Knbtz}. Hence, the positions of ``geon glitches'' are
\begin{widetext}
    \begin{equation}
\tau^{\text{geon}}_n= \ell 
{\sgn}
\! \left(
\tilde \tau_0 + \arctan\sqrt{(K^{\text{geon}}_n)^2 - 1}\right)\arccos \! \left(\frac{K^{\text{geon}}_n\sqrt{1+ \tan^2(\tilde\tau_0)}-\sqrt{(K^{\text{geon}}_n)^2-1} \tan(\tilde\tau_0)}{(K^{\text{geon}}_n)^2+\tan^2(\tilde\tau_0)}\right).
\label{geonglitch_tau0}
\end{equation}
\end{widetext}
 The argument of the arccos in \eqref{geonglitch_tau0} is 
always greater than zero, so geon glitches are always detectable throughout the trajectory,  with
\begin{subequations}
\begin{align}
\lim_{\tau_0 \rightarrow \pi/2} \tau^{\text{geon}}_n &= \frac{\pi}{2} \ell , 
\\
\lim_{\tau_0 \rightarrow -\pi/2} \tau^{\text{geon}}_n &= -\frac{\pi}{2} \ell. 
\end{align}
\end{subequations}

The effect of changing $\tau_0$ is shown in the Penrose-Carter diagrams in Figure~\ref{fig:Penrose_analysis_tau0}. The key new feature is that when $\tau_0<-\tau_H$, so that the detector operates already before emerging from the white hole region, glitches can occur already before the detector enters the black hole region. 

\subsubsection{$t_0 = 0 = \tau_0$}

Consider finally the case in which $t_0 = 0 = \tau_0$. Equation \eqref{geonglitch_tau0} assumes now 
the simple form 
\begin{equation}
\tau^{\text{geon}}_n= \ell\arccos \! \left(\frac{1}{K^{\text{geon}}_n}\right).
\label{geonglitch}
\end{equation}
Since $K^{\text{geon}}_n>q$, we have $\tau^{\text{geon}}_n > \tau_H$, for all~$n$. 
It follows that the distinct topologies of the two black holes, in terms of their signature glitches, only become evident after crossing the horizon.  

\begin{figure}[t]
    \centering
\includegraphics[width=\linewidth]{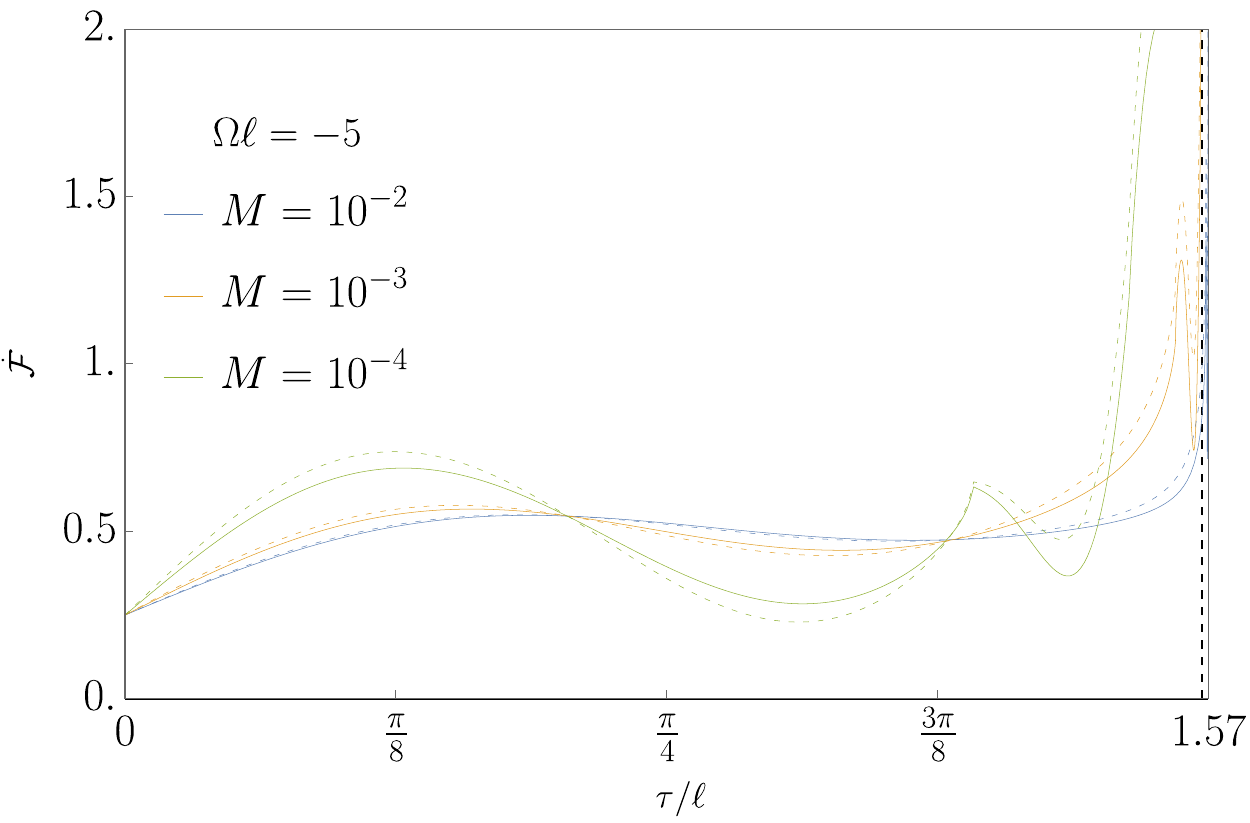}
    \caption{Transition rate for  {$t_0 = 0 = \tau_0$}, $q =100$, $\zeta = 0$, and $\Omega\ell = -5 $ on the BTZ (solid) and geon (dashed) spacetimes for selected values of~$M$. The vertical dashed line is the horizon, located at $\tau/\ell = \arccos(1/q) = 1.5608$ for all values of~$M$.}
    \label{fig:diff_mass}
\end{figure}

\section{Numerical results}
\label{numericalresults}

In this section, we present results from numerical calculations of the transition rate of a detector in each spacetime. Subsection \ref{subsec:t0tau} addresses the special case $t_0=0=\tau_0$, in which case the detector is switched on at the moment of maximum radius on the trajectory, and, on the geon, this moment is at the distinguished $t=0$ hypersurface. 
Subsections \ref{subsec:tnot=tau} and \ref{subsec:t=nottau} address respectively the cases where $t_0 \ne 0=\tau_0$ and $t_0 = 0 \ne \tau_0$. The features present in the generic case, $t_0 \ne 0 \ne \tau_0$, can be deduced from these.

\begin{figure}[t!]
    \centering
     \includegraphics[width=\linewidth]{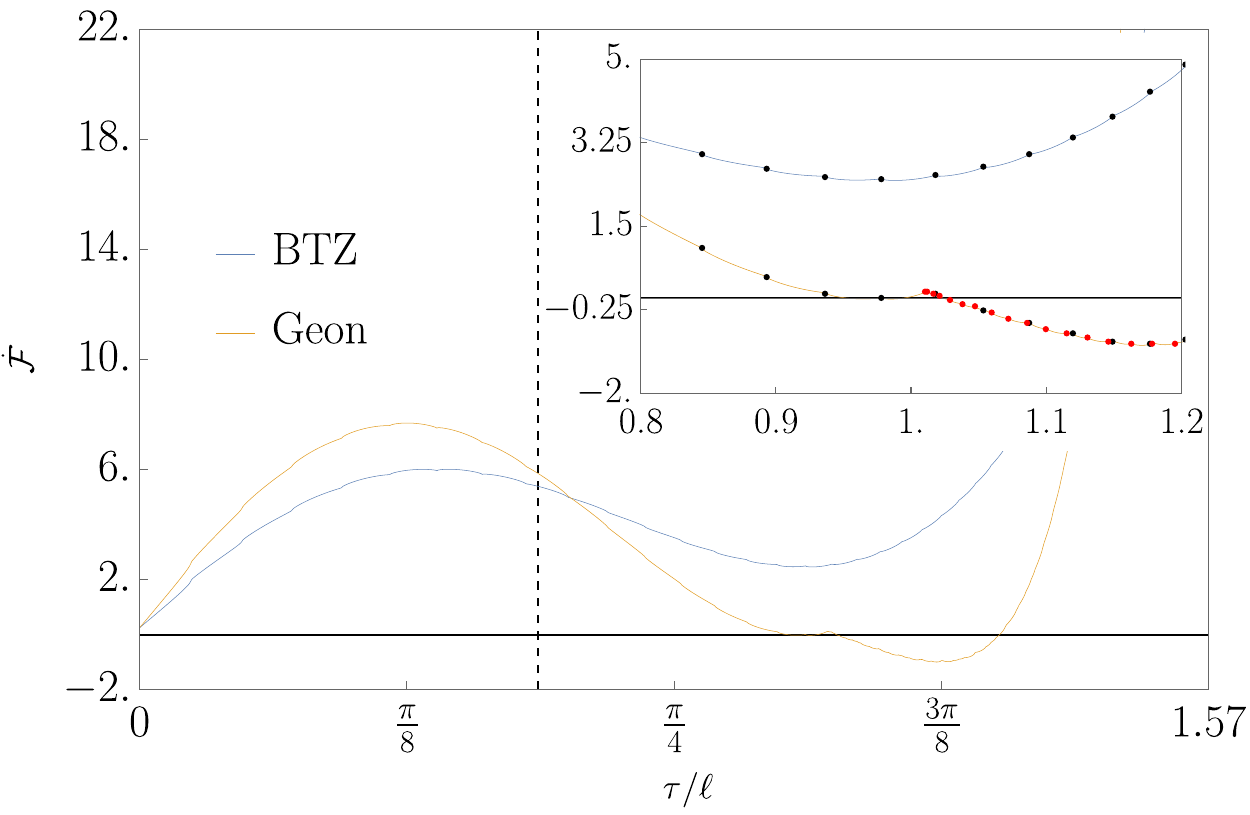}
    \caption{Transition rate for  {$t_0 = 0 = \tau_0$}, $q =1.2$, $\zeta = 0$, $M=10^{-4}$ and $\Omega\ell = -5$ in the BTZ (blue) and geon (orange) spacetimes. The vertical dashed line is the horizon, located
    at $\tau/\ell = \arccos(1/q) = 0.5856$. The black dots represent the glitches associated with the BTZ \eqref{btzglitch} spacetime for both BTZ and geon black holes, while the red dots are the geon glitches~\eqref{geonglitch}; the inset shows detail in the vicinity of the first geon glitch, located inside the horizon.}
    \label{fig:geon_example}
\end{figure}

\begin{figure*}[p!]
    \centering
    \includegraphics[width=0.49\linewidth]{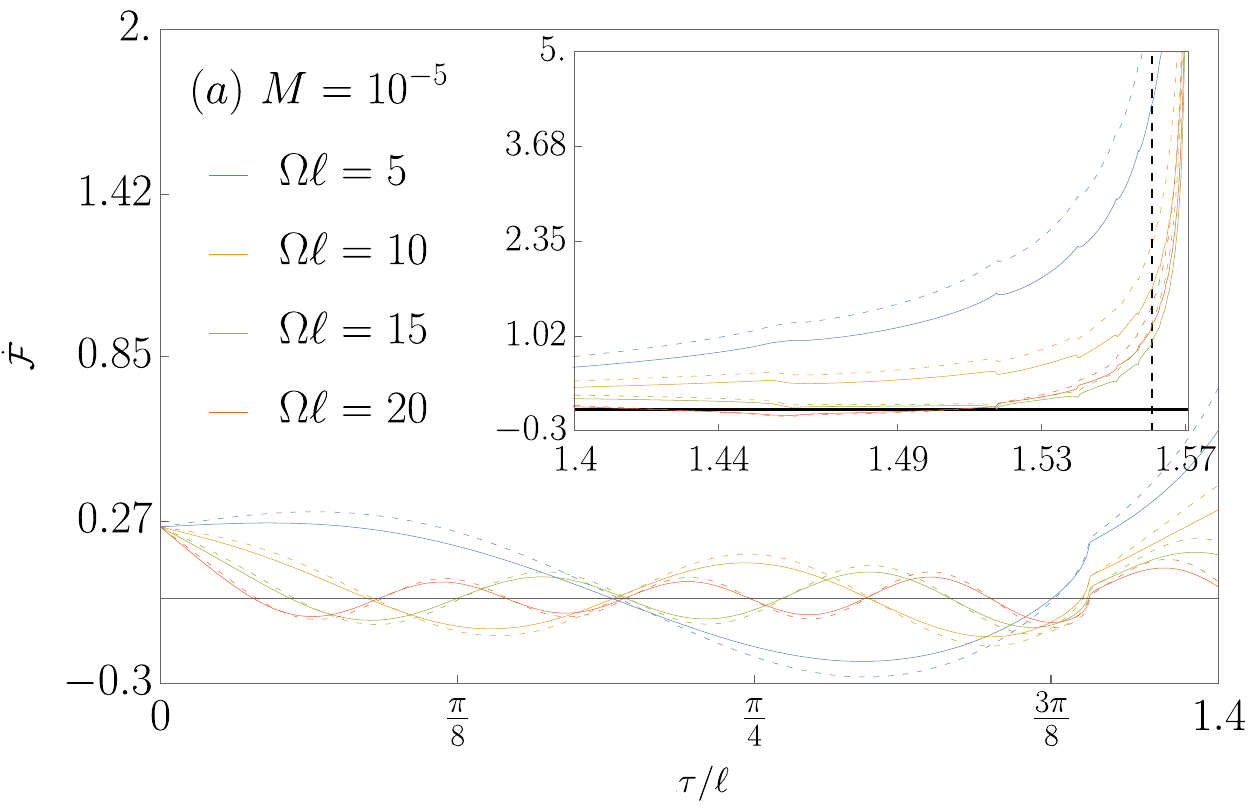}
    \includegraphics[width=0.49\linewidth]{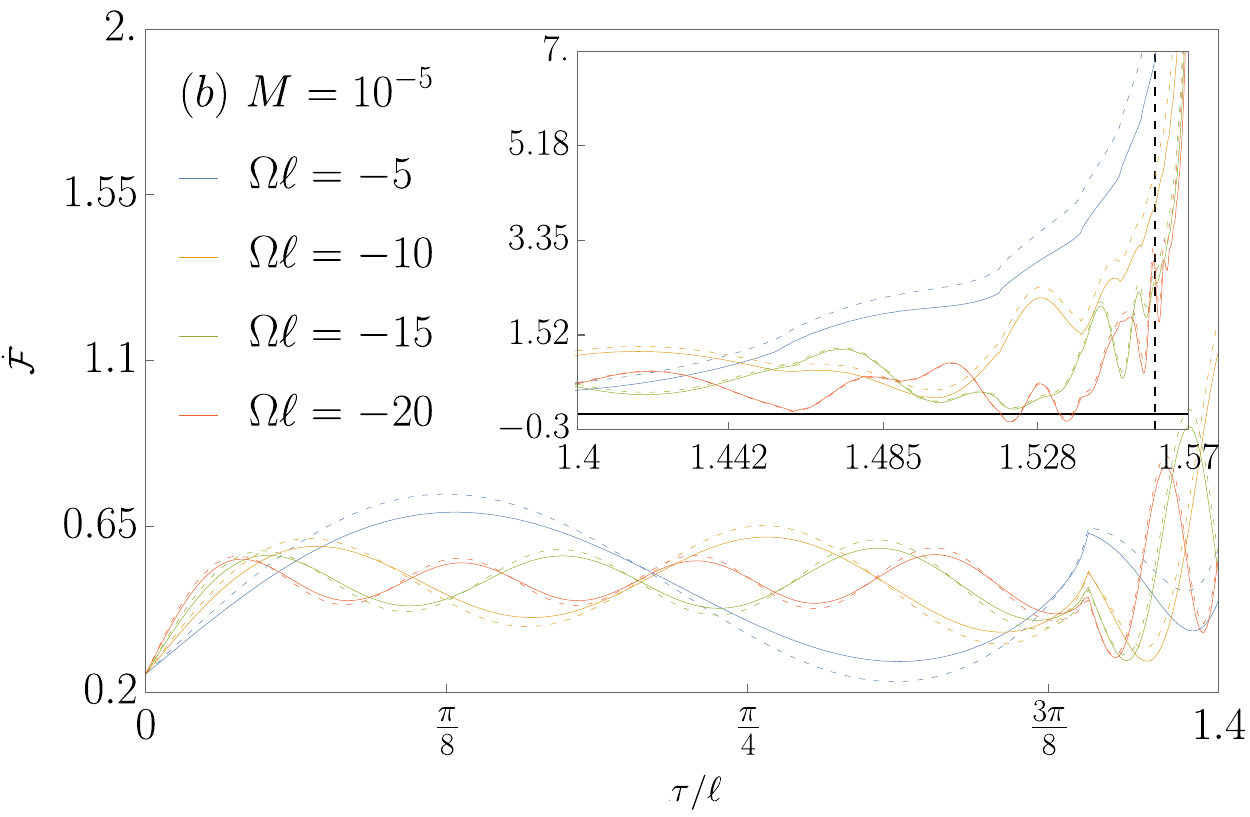}
    \includegraphics[width=0.49\linewidth]{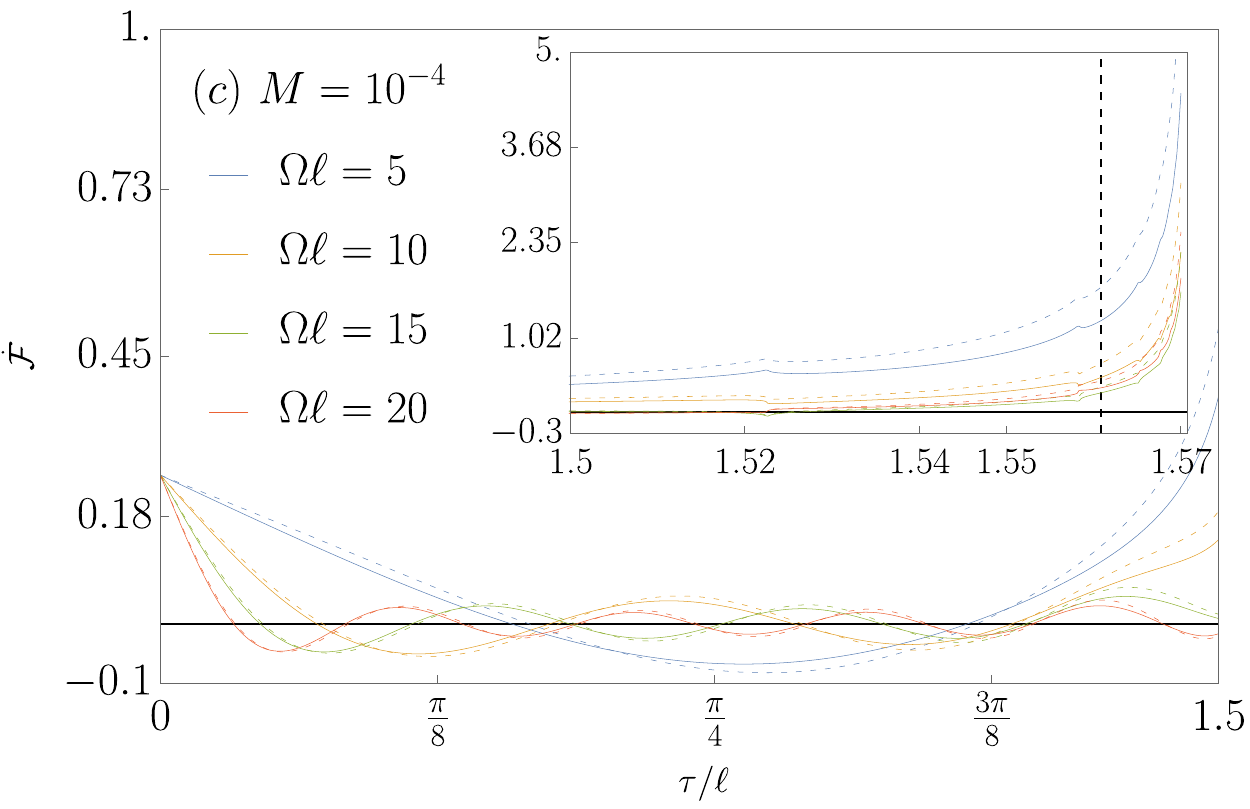}
    \includegraphics[width=0.49\linewidth]{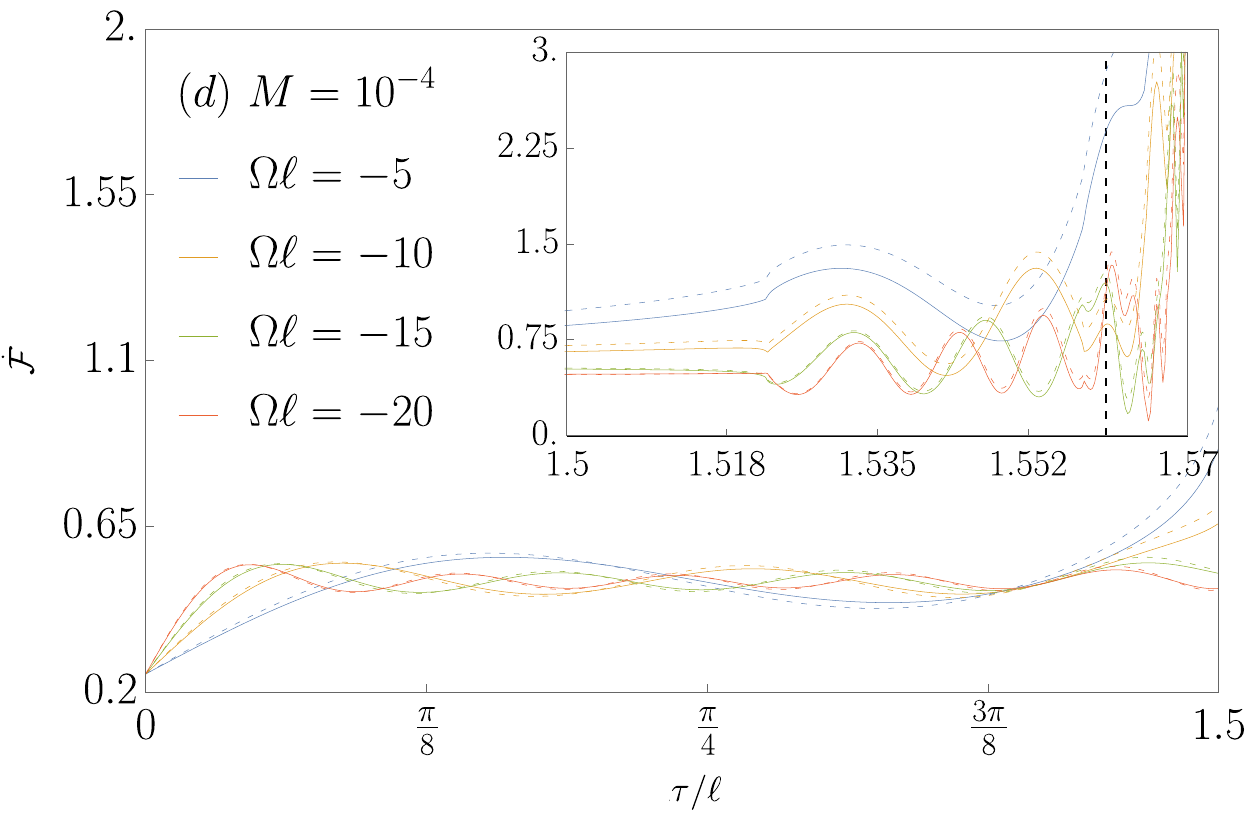}
    \includegraphics[width=0.49\linewidth]{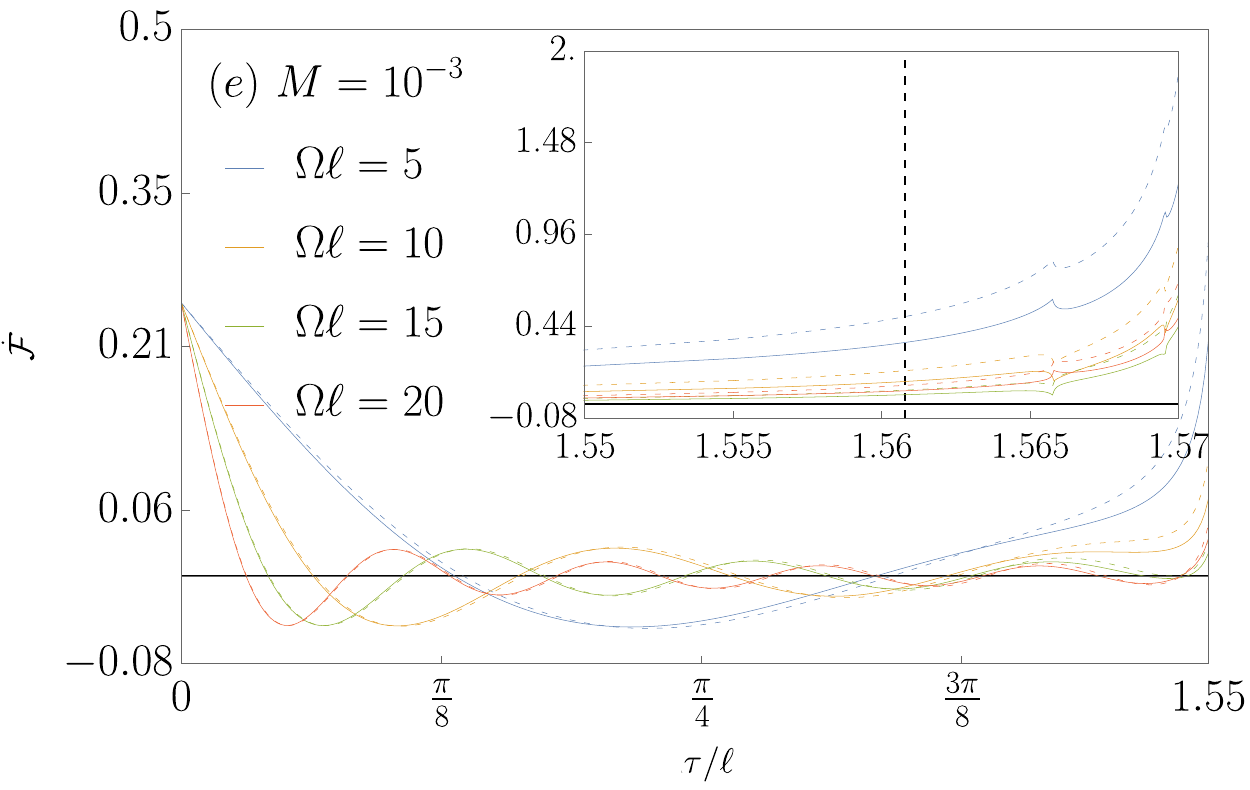}
    \includegraphics[width=0.49\linewidth]{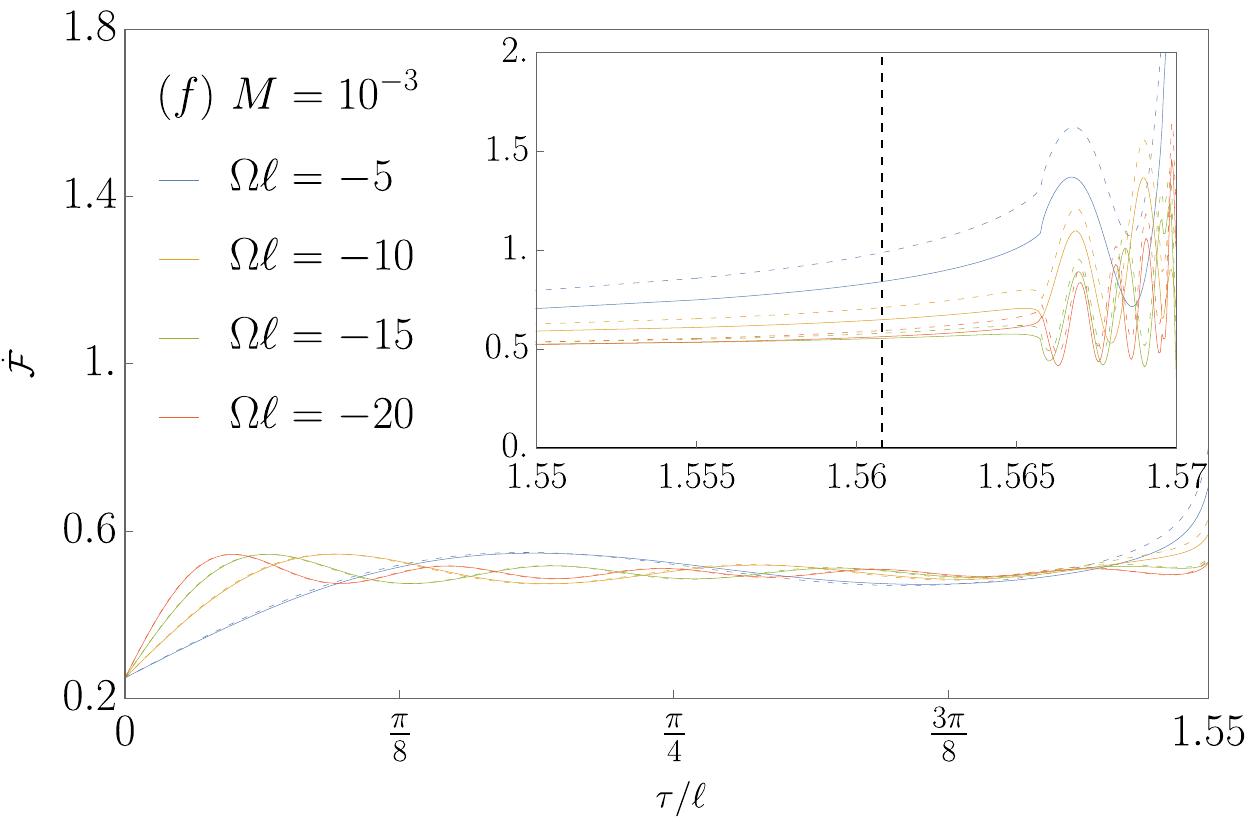}
    \caption{Transition rate for  {$t_0 = 0 = \tau_0$}, $q =100$ and $\zeta = 0$ in the BTZ (solid) and geon (dashed) spacetimes for various positive energy gaps (left column) and negative ones (right column), and various masses $M=10^{-5}$ (top),
    $M=10^{-4}$ (middle) and
    $M=10^{-3}$ (bottom). The vertical dashed line is the horizon; insets show detail in the vicinity of the horizon and the singularity.}
    \label{fig:t0=0,tau_0=0}
\end{figure*}

We set $\zeta=0$ through subsections 
\ref{subsec:t0tau}--\ref{subsec:t=nottau}. We comment on the boundary conditions $\zeta=\pm1$ in Subsection~\ref{subsec:zetanonzero}. 

To ensure convergence of the numerics,  
the number of terms used was calculated so that the first neglected term in the sums \eqref{btz-transition_rate}, \eqref{geon-transition_rate} for every $\tau/\ell \in [0,1.57]$ was less than $\epsilon = 10^{-5}$.

\subsection{$t_0 = 0 = \tau_0$}
\label{subsec:t0tau}

Consider first the case $t_0 = 0 = \tau_0$, shown in Figure \ref{fig:Penrose_analysis_t0} top panel. 

Figure \ref{fig:diff_mass} shows the transition rate as a function of $\tau/\ell$ for $q=100$ and $\Omega\ell = -5$, for selected values of~$M$. 
The difference between the BTZ spacetime and the geon spacetime grows as the mass decreases, with the amplitude 
on the geon being larger. For both, the oscillations increase in both frequency and amplitude further from the horizon as the mass decreases.

Figure \ref{fig:geon_example}
shows the transition rate as a function of $\tau/\ell$ for a detector released much closer to the horizon, at
$q =1.2$, for $M=10^{-4}$. The first geon glitch is clearly discernible in the plot, behind the horizon at ${\tau^{geon}_0}/{\ell} \sim 1.0102$, by \eqref{geonglitch} with $n=0$. 

Figure \ref{fig:t0=0,tau_0=0} shows the transition rate as a function of $\tau/\ell$ for a range of positive and negative energy gaps and a range of masses. 
As with Figure~\ref{fig:geon_example}, it is immediately noticeable that the geon and BTZ curves are not identical even before crossing the horizon. In both cases the rate exhibits quasi-oscillations 
outside the horizon of approximately the same frequency, but with the amplitude of the geon rate being slightly larger. The oscillation frequency decreases with decreasing gap magnitude $|\Omega\ell|$ and increasing mass, as already seen in Figure~\ref{fig:diff_mass}.

This trend persists until the first glitch, after which 
the geon transition rate grows more rapidly than the BTZ rate as the detector approaches the $r=0$ singularity, for both signs of~$\Omega$. 
Outside the horizon, the glitches agree for the two spacetimes, at the positions determined by~\eqref{btzglitch}, as is clearly visible in the upper diagrams where $M=10^{-5}$. 
The geon glitches occur only beyond the horizon 
and are easily visible for the small value of $q$ in Figure \ref{fig:geon_example}. They are not easily visible in Figure \ref{fig:t0=0,tau_0=0}, where the large value of $q$ implies that these glitches are close to the singularity $r=0$, where $\tau \to \pi/2$.

\begin{figure*}[p!]
    \centering
    \includegraphics[width=0.49\textwidth]{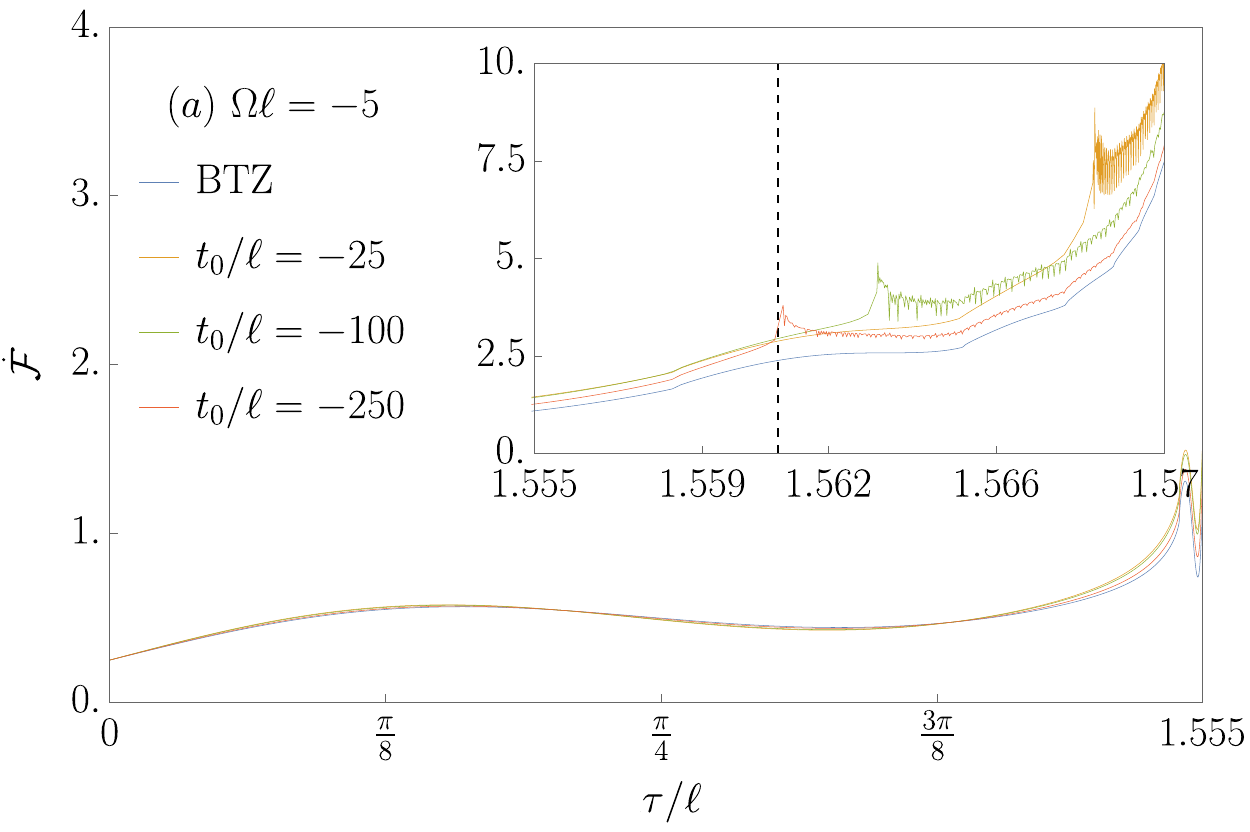}
    \includegraphics[width=0.49\textwidth]{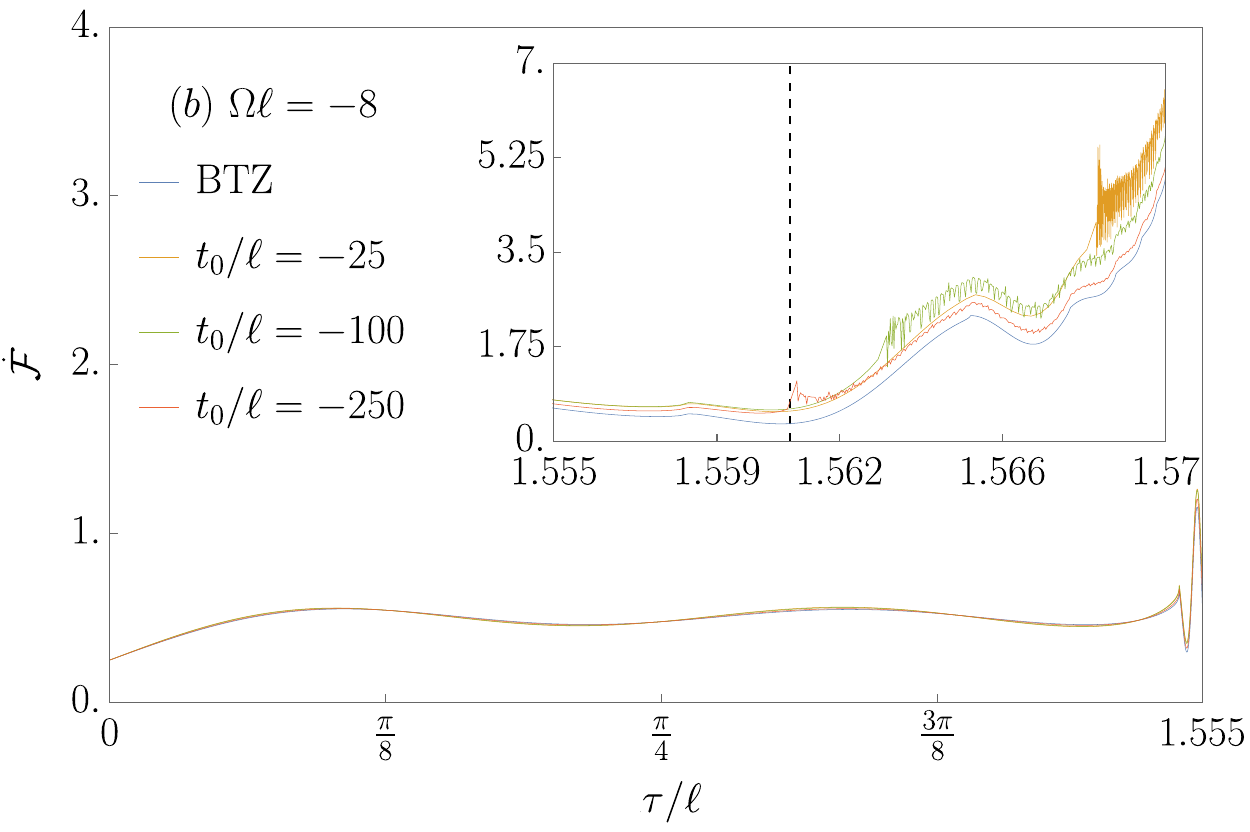}
    \includegraphics[width=0.49\textwidth]{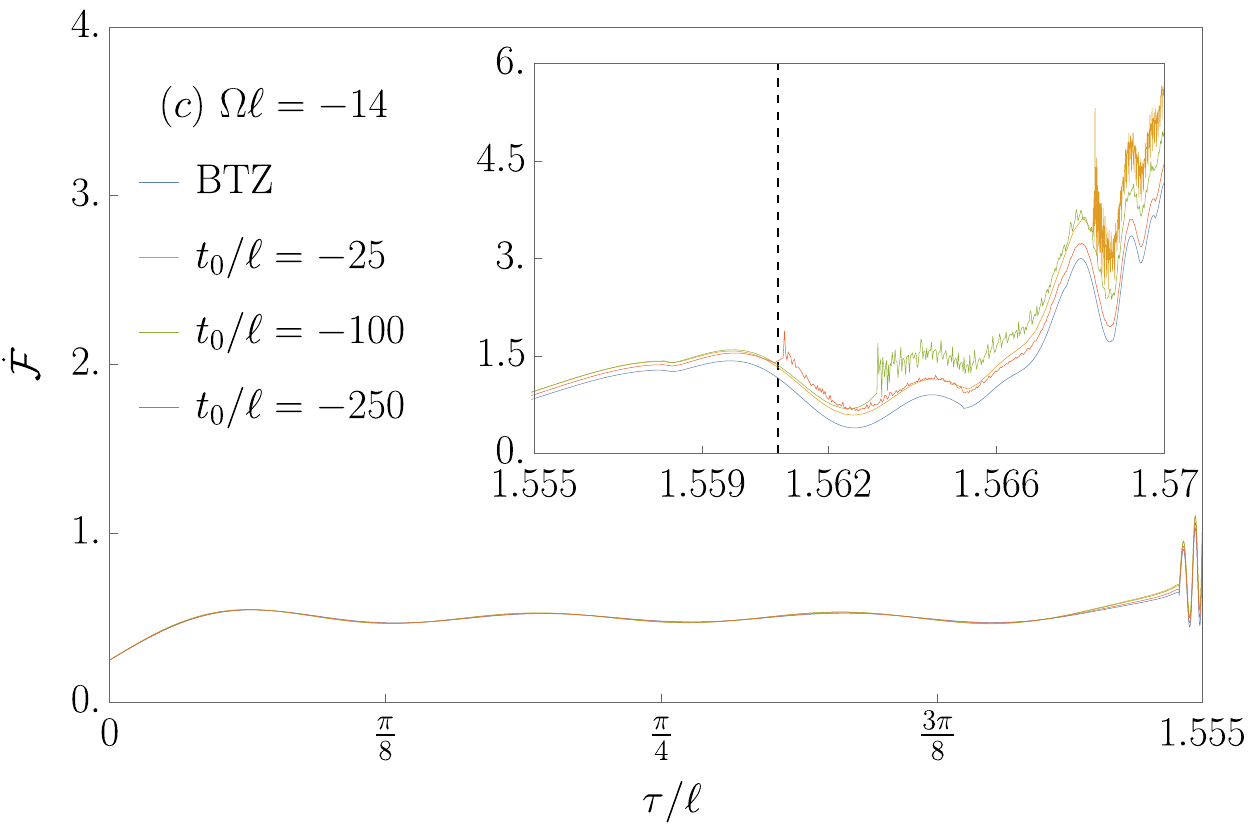}
    \includegraphics[width=0.49\textwidth]{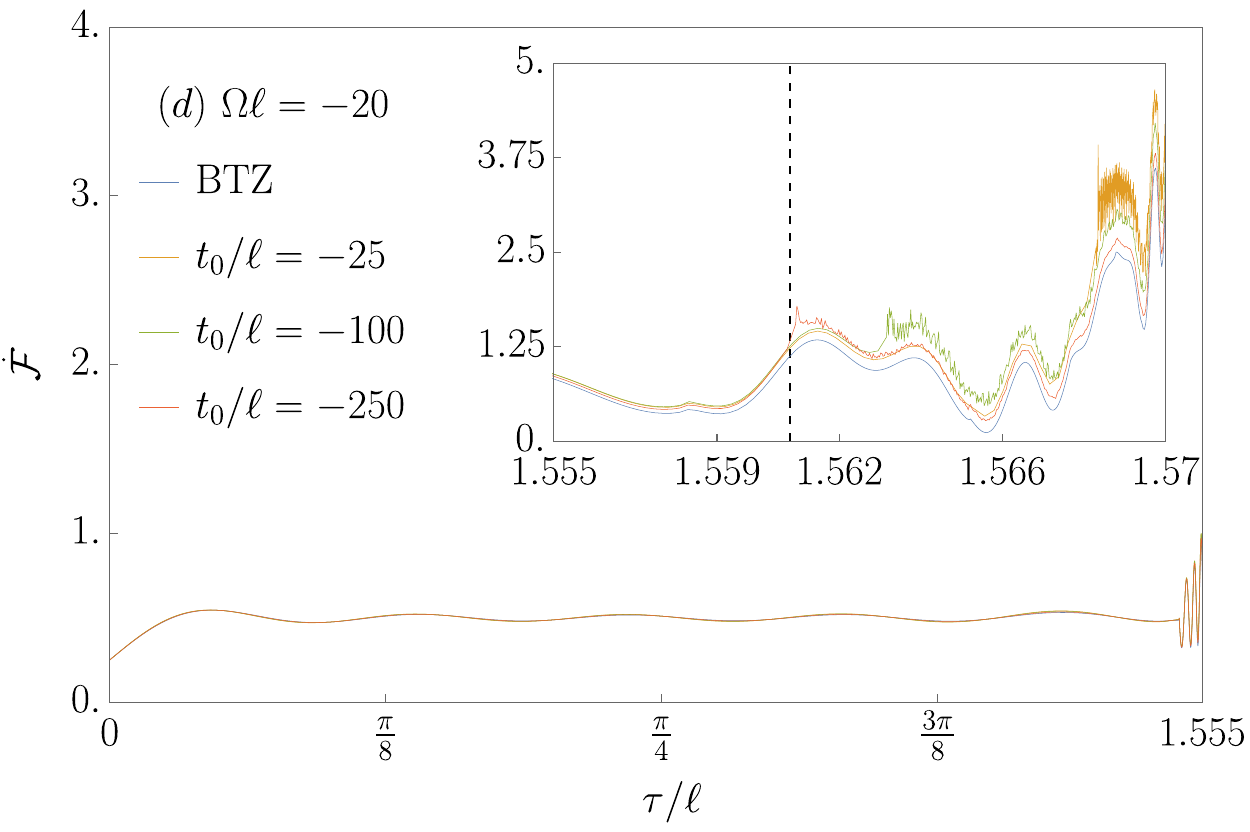}
    \caption{Transition rate for  {$t_0<0 = \tau_0$}, $M = 10^{-4}$, $q =100$, and $\zeta = 0$ for both
    BTZ (blue) and geon spacetimes for three different values of $t_0 < 0$. The BTZ rate is the same for each $t_0$ but the geon rate is not, with the differences becoming noticeable near and beyond the horizon (vertical dashed line), as shown in the insets.  Each panel depicts a different choice of gap $\Omega$.
    }
    \label{fig:tnoughtneg}
\end{figure*}

The transition rate is relatively insensitive to the value of the energy gap $\Omega$ beyond the first glitch. The effect of the gap is mainly observed in the BTZ case~\eqref{btz-transition_rate}, where  for positive gaps there are oscillations after the first glitch that increase with $\Omega$~\cite{MariaRosaBTZ}. For negative gaps, the trend remains monotonically increasing, interspersed with glitches.

What seems to particularly influence the transition rate is the mass~$M$. As the mass increases, two effects are observed: the glitches  move to the right, and there is a noticeable decrease in $\Delta\dot{\mathcal{F}}_\tau$.  
This effect can be easily observed by examining the expressions 
\eqref{WBTZ-and-geon}, 
where it is clear from \eqref{sigma_BTZandgeon}
that for $M \rightarrow \infty$, the only surviving term is the zeroth-order term of (\ref{WBTZ}). The cause of this effect is the dependence of $K^{\text{BTZ}}_n$ and $K^{\text{geon}}_n$ on~$M$. 
However, since $K^{\text{geon}}_n > K^{\text{BTZ}}_n$ for fixed $n$, $M$, and $q$, the terms in the sum related to the geon are suppressed more rapidly.

\subsection{$t_0 \neq 0 = \tau_0$}
\label{subsec:tnot=tau}

Consider next the case $t_0 \neq 0 = \tau_0$, shown in Figure \ref{fig:Penrose_analysis_t0} middle and bottom panels. As discussed in Section~\ref{analyticresults}, the value of $t_0$ does not affect the transition rate on the BTZ spacetime, but it does affect the transition rate on the geon. 

Figure~\ref{fig:tnoughtneg} shows the transition rate on the BTZ spacetime and the geon spacetime with three different negative values of $t_0$ and various values of~$\Omega$, the other parameters being $M=10^{-4}$, $q=100$ and $\zeta=0$. The differences between the BTZ transition rate and the geon transition rate again become significant only when the detector is close to the horizon, and when $t_0$ is large and negative, this happens very close to the horizon-crossing moment in the detector's proper time, because the detector then spends little proper time between the distinguished Killing time surface and the horizon, as is seen from the bottom panel in Figure~\ref{fig:Penrose_analysis_t0}. The jitter persisting after the first glitch is caused by the subsequent glitches.

\begin{figure}[t!]
    \centering
     \includegraphics[width=\linewidth]{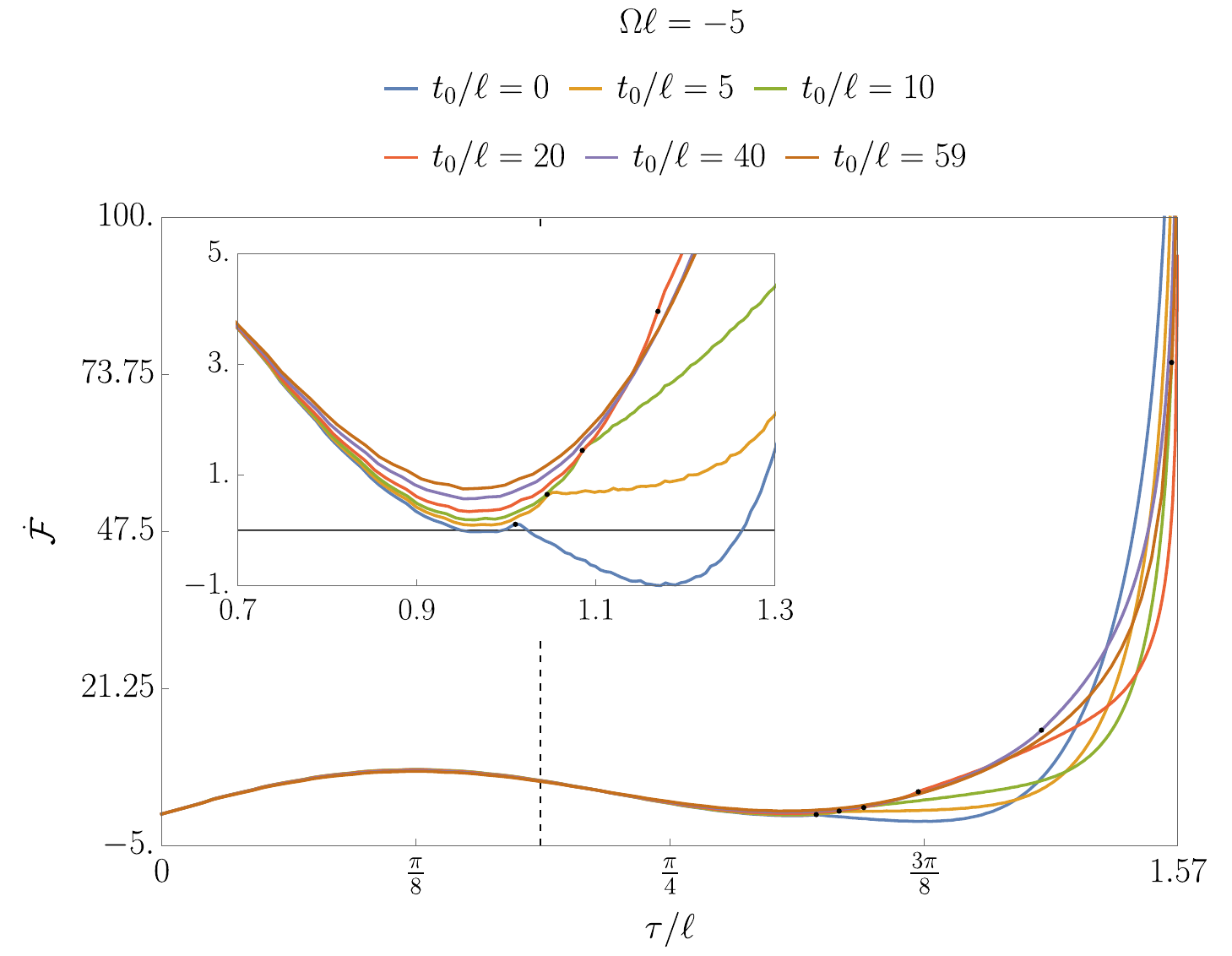}
\caption{Transition rate for  {$\tau_0 = 0 \le t_0$}, 
$q =1.2$, $\zeta = 0$, $M=10^{-4}$ and $\Omega\ell = -5 $ in the geon spacetimes. The vertical dashed line is the horizon, located at $\tau/\ell = \arccos(1/q) = 0.5856$. 
The black dots represent the geon glitches, given by~\eqref{geonglitch_t0}. 
These glitches move rightward as $t_0$ increases towards $t_0^{lim} \approx 59.947 \ell$ \eqref{tlim}, and there are no glitches for 
larger values of~$t_0$.
The inset shows detail in the vicinity of the first geon glitch for each $t_0/\ell$.}
    \label{fig:geon_t0glitch_example}
\end{figure}

\begin{figure}[t!]
    \centering
     \includegraphics[width=\linewidth]{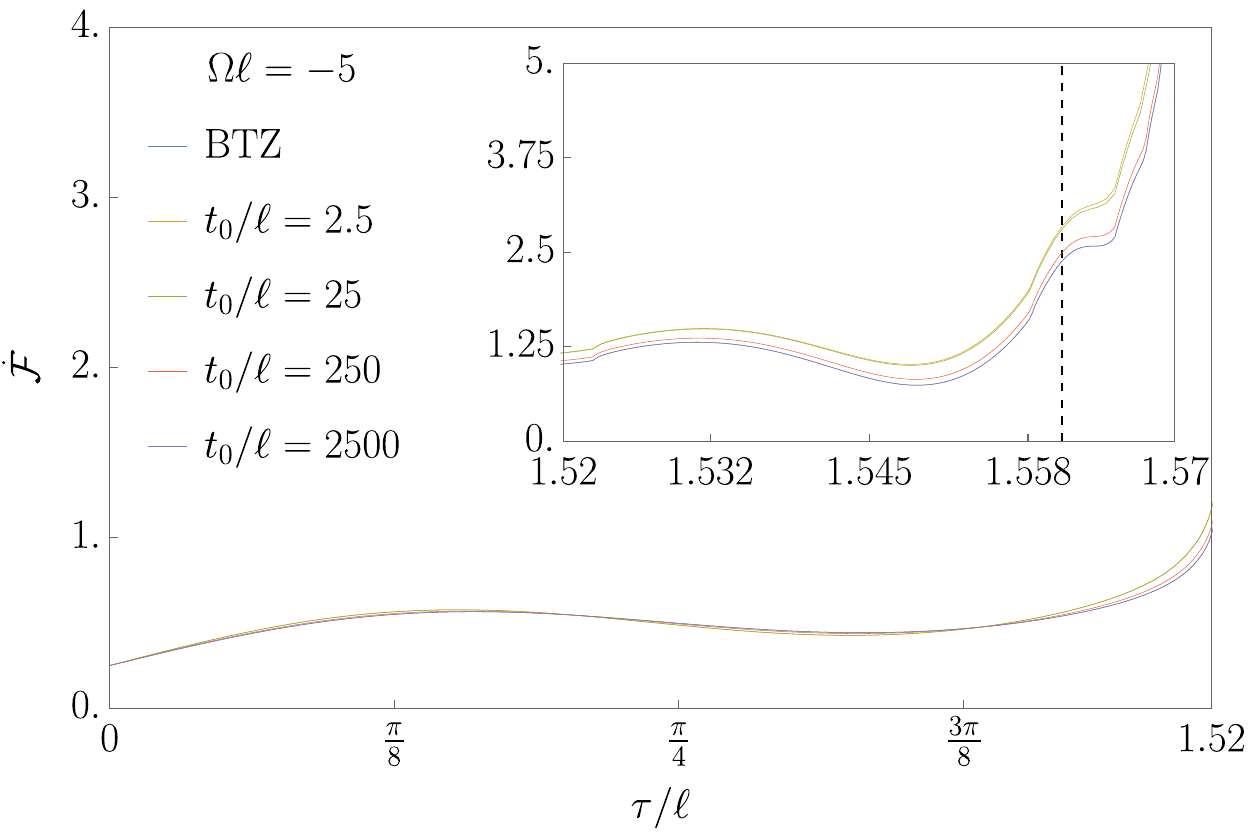}
    \caption{Transition rate for {$\tau_0 = 0 < t_0$}, $q =100$, $\zeta = 0$, $M=10^{-4}$ and $\Omega\ell = -5$ in the geon spacetimes. The vertical dashed line is the horizon, located
    at $\tau/\ell = \arccos(1/q) = 1.5608$. The inset shows detail in the vicinity of the horizon for each $t_0/\ell$.}
    \label{fig:geon_t0_example}
\end{figure}

Figures \ref{fig:geon_t0_example} and \ref{fig:geon_t0glitch_example} shows the transition rate on the geon spacetime for various positive values of $t_0$ and selected values of the other parameters. The glitch structure behind the horizon is now suppressed for large positive~$t_0$, as was to be expected from the middle panel in Figure~\ref{fig:Penrose_analysis_t0}.

\begin{figure}[t!]
    \centering
     \includegraphics[width=\linewidth]{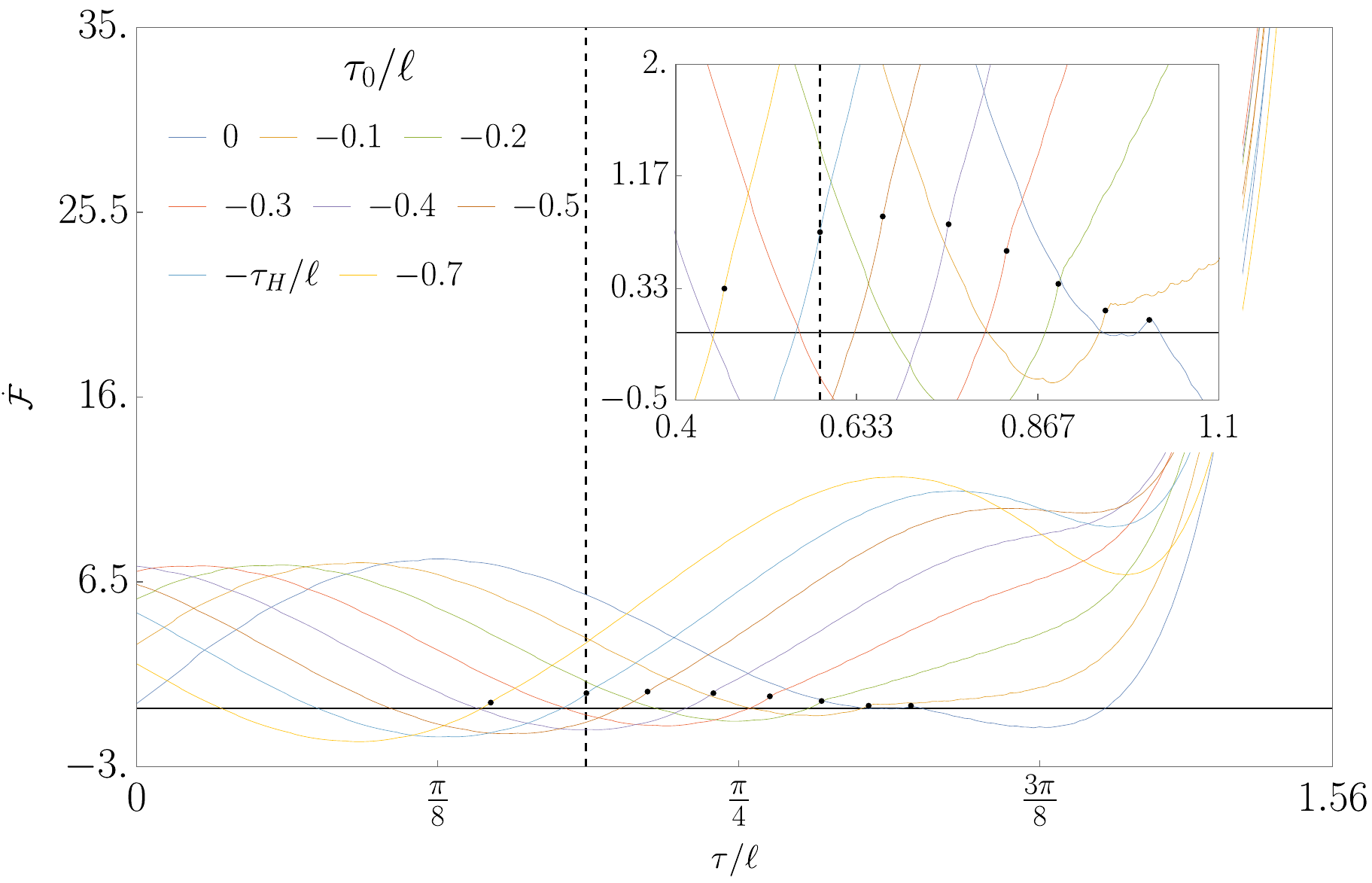}
    \caption{Transition rate for  {$\tau_0 \le 0 = t_0$}, $q =1.2$, $\zeta = 0$, $M=10^{-4}$ and $\Omega\ell = -5 $ in the geon spacetimes. The vertical dashed line is the horizon, located
    at $\tau/\ell = \arccos(1/q) = 0.5856$. The black dots represent the glitches associated with the geon spacetime \eqref{geonglitch_tau0};  the inset shows detail in the vicinity of the first geon glitch for each $\tau_0/\ell$.}
    \label{fig:geon_tau0_example}
\end{figure}

\begin{figure*}[h!]
    \centering
    \includegraphics[width=0.49\textwidth]{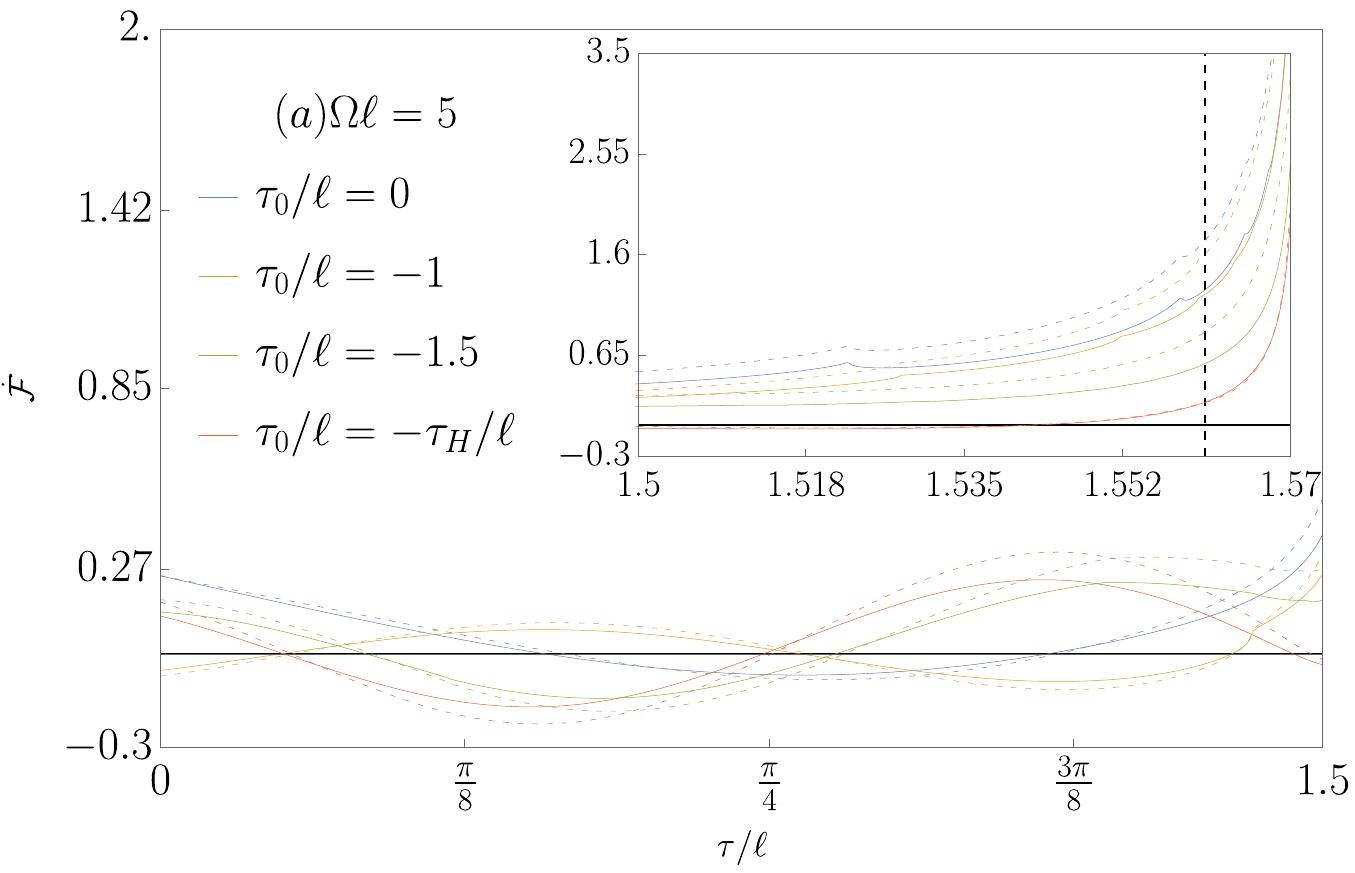}
    \includegraphics[width=0.49\textwidth]{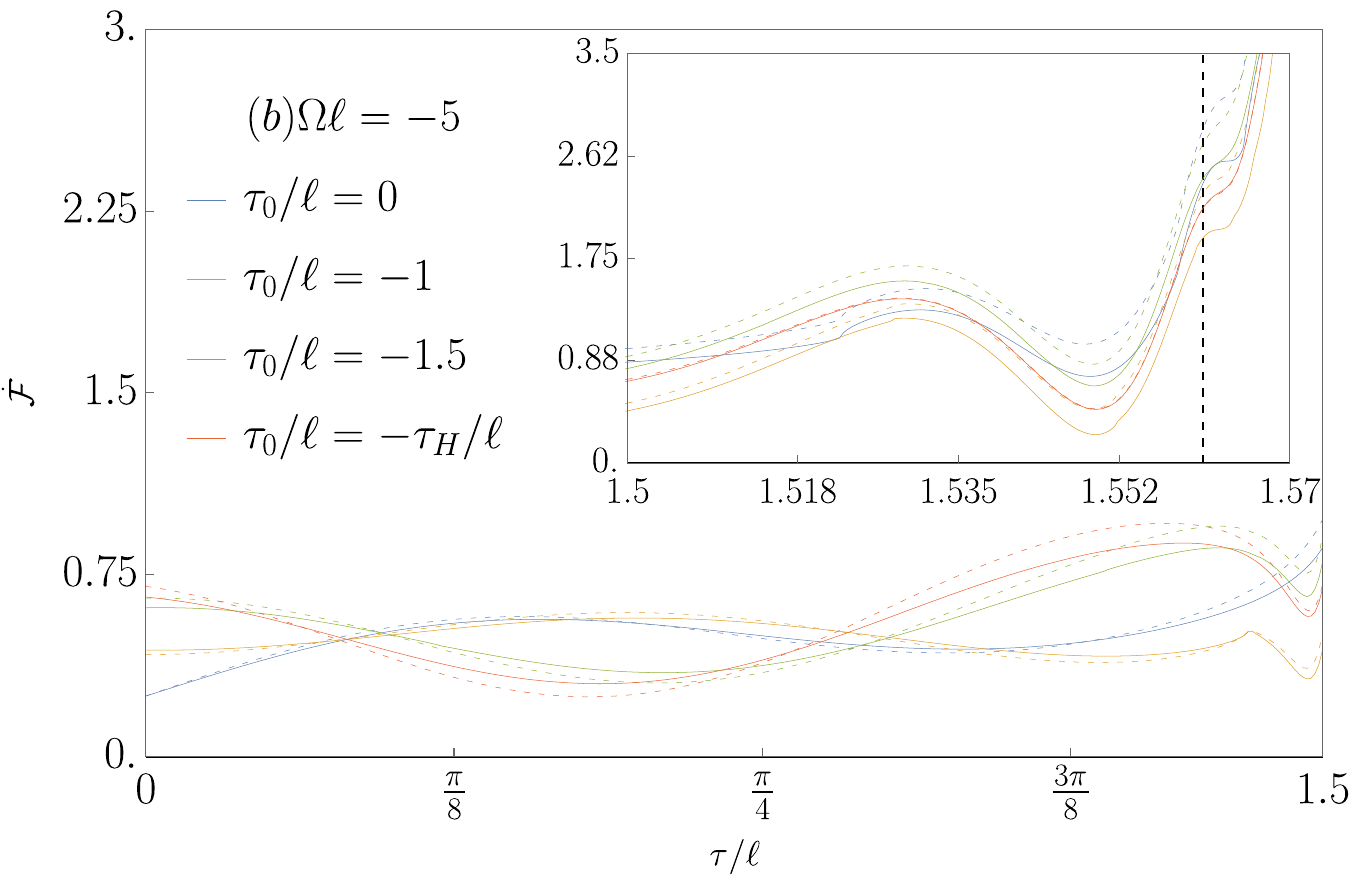}
    \caption{
    Transition rate for  {$\tau_0 \le 0 = t_0$}, $M = 10^{-4}$, $q =100$, and $\zeta = 0$ for both BTZ (solid) and geon (dashed) spacetimes for differing values of $\tau_0 < 0$ and positive
    (left) and negative (right) energy gaps. Insets illustrate the behaviour near and inside the horizon, shown as a vertical dashed line.} 
    \label{fig:taunoughtneg}
\end{figure*}

\begin{figure*}[h!]
    \centering
    \includegraphics[width=0.49\textwidth]{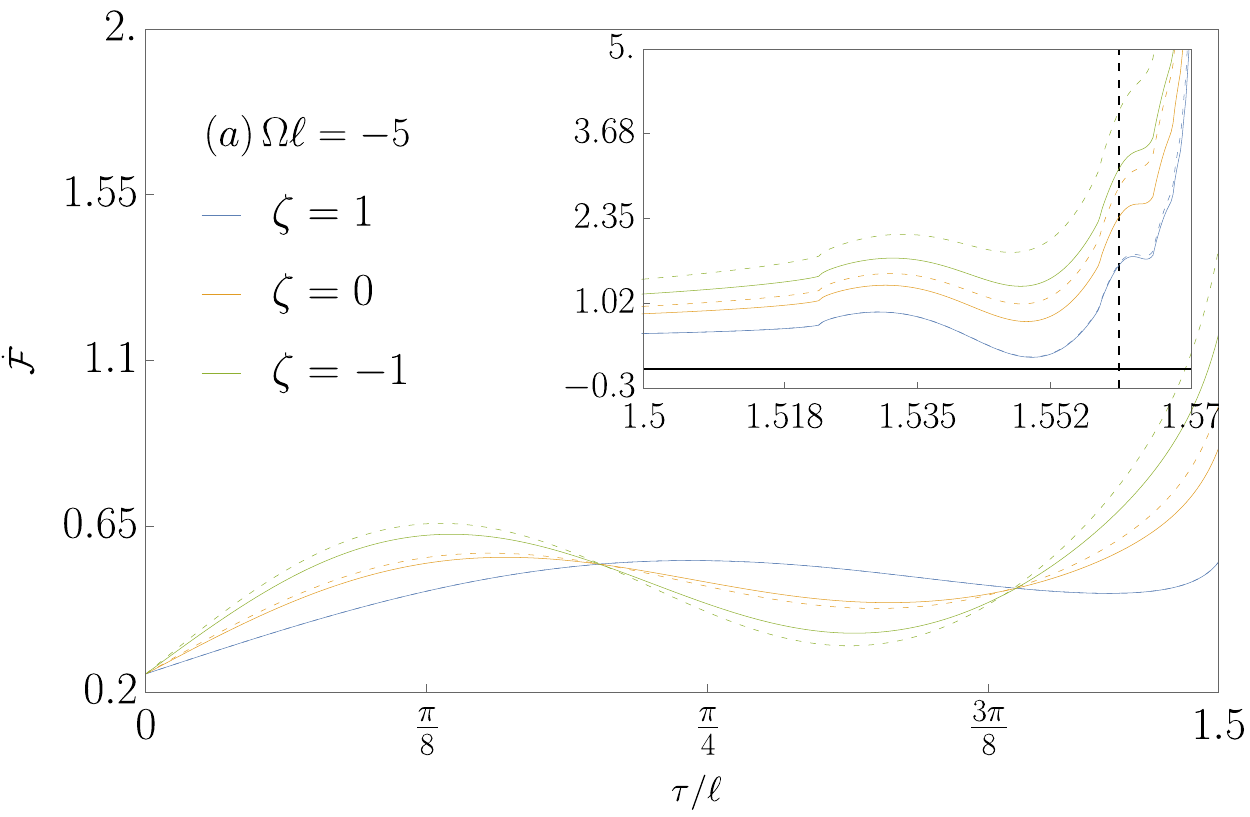}
    \includegraphics[width=0.49\textwidth]{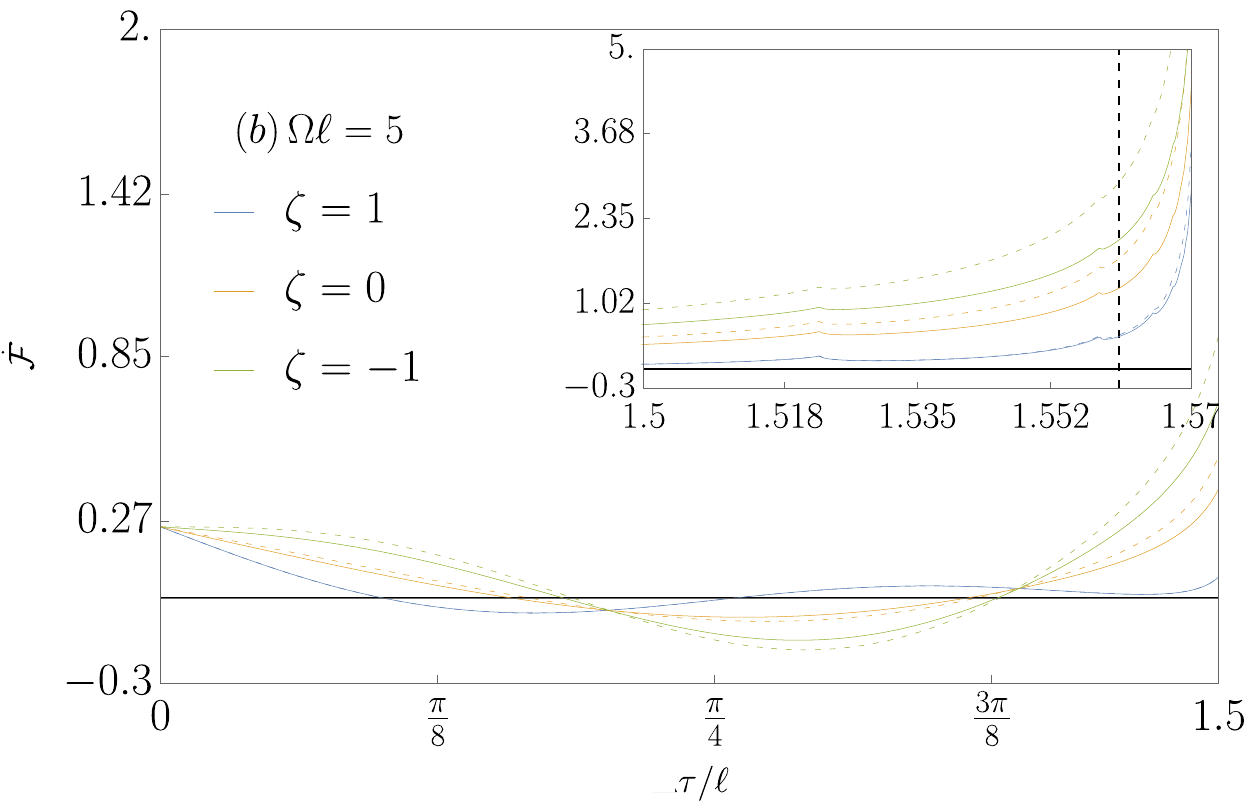}
    \caption{
    Transition rate for $t_0 = 0 = \tau_0$, $M = 10^{-4}$ and $q =100$ for both BTZ (solid) and geon (dashed) spacetimes for differing values of $\zeta \in \{1,0,-1\}$ and positive
    (left) and negative (right) energy gaps. Insets illustrate the behaviour near and inside the horizon, shown as a vertical dashed line. 
     }
    \label{fig:diff-zeta}
\end{figure*}

\begin{figure}[h!]
    \centering
     \includegraphics[width=\linewidth]{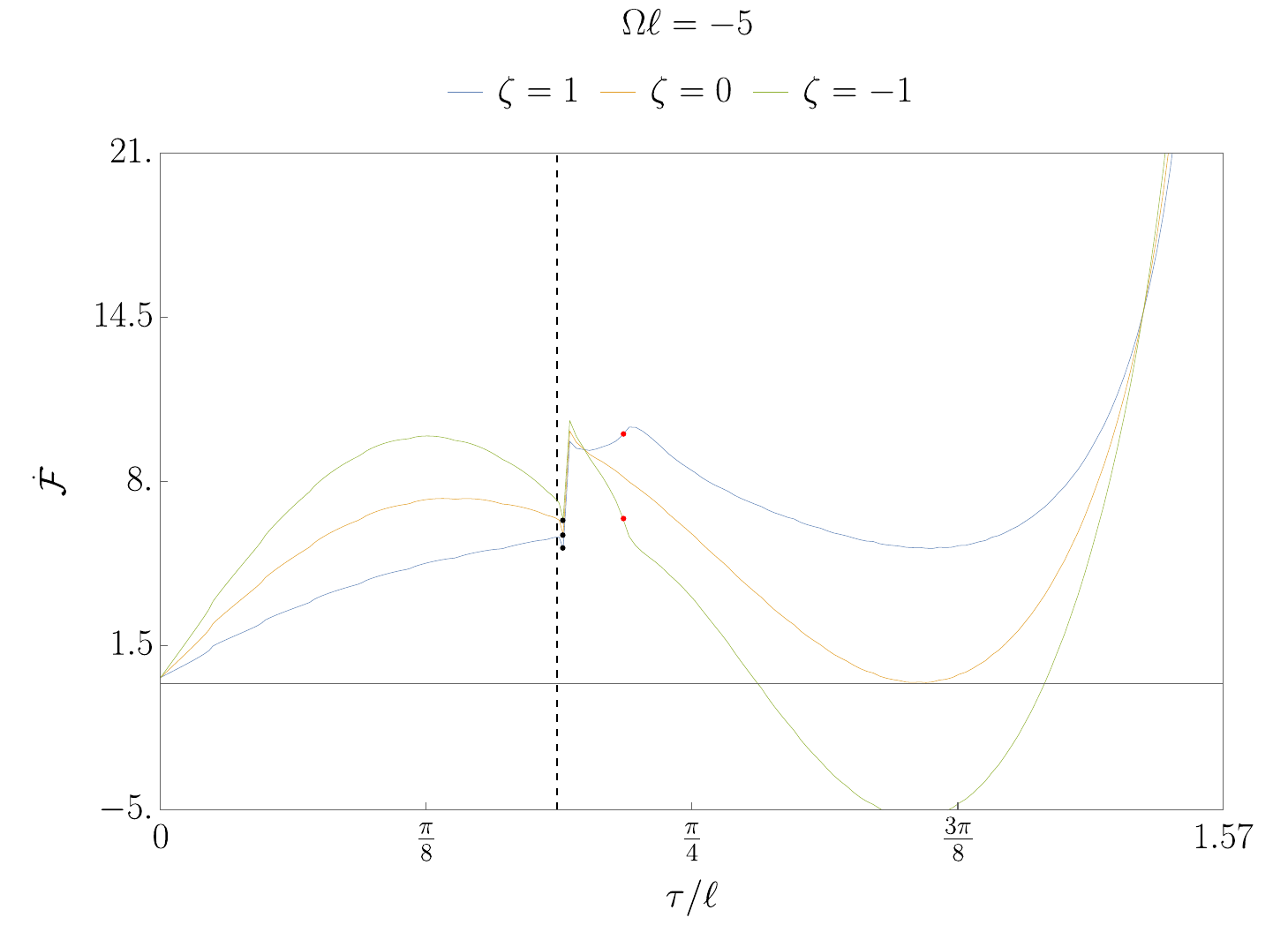}
    \caption{Transition rate in the geon spacetimes for $\tau_0=0$, $t_0/\ell = -200$,  $q =1.2$, $M=10^{-4}$, $\Omega\ell = -5 $ and $\zeta \in \{1,0,-1\}$. The vertical dashed line is the horizon, located
    at $\tau/\ell = \arccos(1/q) = 0.5856$.  {The black dots are the glitches from the $\zeta$-independent term, at~\eqref{geonglitch_t0}. The red dots are the glitches from the term proportional to~$\zeta$, at~\eqref{geonglitch_boundary}; as discussed around~\eqref{tlim_boundary}, these are present only for $t_0/\ell < t_0^{\text{b},\text{lim}}/\ell \approx -59.947$.}}
    \label{fig:geon_t0_boundary}
\end{figure}

\subsection{$t_0 = 0 \ne \tau_0$}
\label{subsec:t=nottau}

Consider finally the case $t_0 = 0 \ne \tau_0$. 

We consider only the case $\tau_0<0$, in which the detector starts to operate before reaching the maximum value of $r$ on the distinguished $t=0$ hypersurface. This is the spacetime diagram in Figure~\ref{fig:Penrose_analysis_tau0}. 
Numerical results are shown in Figures \ref{fig:geon_tau0_example} and~\ref{fig:taunoughtneg}. 

The figures show that glitches move toward earlier proper times as $\tau_0$ becomes more negative. 
Glitches before crossing the black hole horizon can occur only when $\tau_0 < -\tau_H$, so that the detector has started to operate before emerging from the white hole part of the spacetime, as shown in Figure \ref{fig:Penrose_analysis_tau0} lower panel: one such glitch is seen for one of the trajectories in Figure~\ref{fig:geon_tau0_example}. 

\subsection{$\zeta\ne0$}
\label{subsec:zetanonzero}

We now comment on what differs for the Dirichlet and Neumann boundary conditions, $\zeta=\pm1$. 

Over most of the parameter space, we find that the $\zeta=\pm1$ response is qualitatively similar to the $\zeta=0$ response. A sample plot for $t_0 = 0 = \tau_0$ is shown in Figure~\ref{fig:diff-zeta}. 

However, there are regions of the parameter space where the $\zeta=\pm1$ term introduces new glitches in the response. Geometrically, these new glitches come from null geodesics that connect the switch-on moment to a later moment on the trajectory by reflection from infinity. Consideration of Figure \ref{fig:Penrose_analysis_t0} lower panel shows that one situation where this happens is for  $t_0<0=\tau_0$ when $t_0$ is sufficiently negative, and the new glitches then occur after the detector has entered the black hole. Analytically, the new glitches come by solving 
$\sigma_n^{\text{geon}}\left(\tau^{\text{b}}_n,0\right)+2 =0$, which yields a new set of glitches positioned at 
\begin{equation}
\tau^{\text{b}}_n(t_0) = \pi \ell - \tau^{\text{geon}}_n(-t_0) , 
\label{geonglitch_boundary}
\end{equation}
 where the superscript `b' denotes that these glitches originate from the $\zeta$ term in
\eqref{geon-transition_rate}.
These glitches exist for $t_0<t_0^{\text{b},\text{lim}}$, where 
\begin{equation}
    t_0^{\text{b},\text{lim}}=- \frac{\ell}{2\sqrt{M}} \arccosh \! \left({\frac{q}{\sqrt{q^2-1}}}\right). 
    \label{tlim_boundary}
\end{equation}
All of these glitches occur after the detector has crossed the future horizon, and as $t_0\to-\infty$, they occur shortly after the horizon-crossing, 
\begin{equation}
    \lim_{t_0 \rightarrow -\infty} \tau^{\text{b}}_n = \tau_H   \; .
\end{equation}
For high values of $q$, these glitches are not easily visually distinguishable from the $\zeta=0$ glitches, but Figure \ref{fig:geon_t0_boundary} shows a plot where they are distinguishable for $q=1.2$.

\section{Conclusions}
\label{conc}

We have computed and compared the transition rates of a radially infalling Unruh-DeWitt detector into a BTZ black hole and  the $\mathbb{R}\text{P}^{2}$ geon, the latter being obtained from the former by a topological identification. The geon topology is hidden behind its horizon, and so the two spacetimes are classically indistinguishable outside the horizon.  

Our results are commensurate with previous work on static detector response and entanglement in geon spacetimes
\cite{Smith_2014,Smith_thesis,Henderson:2022oyd}, but reveal additional features.
The pointwise difference between the two rates is indeed different from zero even outside the black hole,
as for the static case \cite{Smith_2014}.  Glitches
observed in the BTZ case
\cite{MariaRosaBTZ} occur in identical places  for the geon
outside of the horizon. 
However there are extra glitches associated with the geon, 
representing a significant difference between the two rates. 
 Assuming that the detector's interaction with the field begins when the detector is in the black hole exterior, these extra glitches occur only after the detector has entered the black hole. 
This result is consistent with the topological censorship theorem \cite{Friedman_1995} and with what one would expect from classical  considerations.

Nevertheless, the infalling detector can probe interior topology prior to encountering the horizon insofar as the amplitude of the geon response rate is larger than that of its BTZ counterpart.  While less dramatic than the signature glitches inside the horizon,  it is significant that an infalling  quantum  detector is sensitive to hidden topology.

\bigskip

\section*{Acknowledgements}

We thank Chris Shallue and Sean Carroll for sharing with us an early copy of their work on the Schwarzschild infall~\cite{Shallue:2025zto}.
We thank an anonymous referee for helpful presentational suggestions. 
This work was supported in part by the Natural Sciences and Engineering Research Council of Canada.  MRPR gratefully acknowledges the support provided by the Mike and Ophelia Lazaridis Graduate Fellowship.
The work of JL was supported by United Kingdom Research and Innovation Science and Technology Facilities Council [grant numbers ST/S002227/1, ST/T006900/1 and ST/Y004523/1]. 
For the purpose of open access, the authors have applied a CC BY public copyright licence to any Author Accepted Manuscript version arising.

\bibliography{ref}

\end{document}